\documentclass[hyper,11pt,paper]{article}
\usepackage[centertags]{amsmath}
\usepackage{amssymb}
\usepackage{graphicx}
\usepackage{epsfig}
\usepackage{ulem}
\usepackage[english]{babel}
\usepackage{array}
\usepackage{amsthm}
\usepackage{latexsym}
\usepackage[mathcal]{euscript}
\pdfoutput=1
\usepackage{epsfig}
 \usepackage{jheppub}
\usepackage{mathdots}
\usepackage{MnSymbol}
\usepackage{physics}
\usepackage{multirow}
\usepackage{ dsfont }
\usepackage{capt-of}
\usepackage{caption}
\usepackage{subcaption}
\usepackage{soul}
\usepackage{comment}

\usepackage{tikz}
\usetikzlibrary{calc,decorations.pathmorphing,arrows.meta,fadings,decorations.markings}
\tikzset{>={Stealth[width=2mm,length=2mm]}}

\newcommand{\cD}{\mathcal{D}}
\newcommand{\cH}{\mathcal{H}}

\newcommand{\cF}{\mathcal{F}}

\newcommand{\bracket}[2]{\langle #1 | #2 \rangle}
\newcommand{\ZZ}{{\mathbb{Z}}}

\newcommand{\Mod}[1]{\ (\mathrm{mod}\ #1)}

\definecolor{dgreen}{rgb}{0.,0.6,0.}

\usepackage{scalerel,stackengine}

\newcommand{\be}{\begin{equation}}
\newcommand{\ee}{\end{equation}}
\newcommand{\ba}{\begin{eqnarray}}
\newcommand{\ea}{\end{eqnarray}}

\title{Holographic dual of Bures metric and subregion complexity}

\author[a]{Marius Gerbershagen,}
\author[a]{Juan Hernandez,}
\author[a,b]{Mikhail Khramtsov,}
\author[a]{and Maria Knysh}
\affiliation[a]{Theoretische Natuurkunde, Vrije Universiteit Brussel (VUB) and The International Solvay Institutes, Pleinlaan 2, B-1050 Brussels, Belgium}
\affiliation[b]{David Rittenhouse Laboratory, University of Pennsylvania,\\
209 S. 33rd Street, Philadelphia, PA 19104, USA}
\emailAdd{marius.gerbershagen@vub.be}
\emailAdd{juan.hernandez@vub.be}
\emailAdd{mikhail.khramtsov311@gmail.com}
\emailAdd{maria.knysh@vub.be}
\abstract{Within the AdS/CFT correspondence, computational complexity for reduced density matrices of holographic conformal field theories has been conjectured to be related to certain geometric observables in the dual gravity theory. We study this conjecture from both the gravity and field theory point of view. Specifically, we consider a measure of complexity associated to the Bures metric on the space of density matrices. We compute this complexity measure for mixed states associated to single intervals in descendant states of the vacuum in 2d CFTs. Moreover, we derive from first principles a geometric observable dual to the Bures metric which is localized in the entanglement wedge of the AdS spacetime associated to the quantum circuit on the boundary. Finally, we compare the Bures metric complexity measure with holographic subregion complexity within the ``complexity=volume'' paradigm for perturbatively small transformations of the vacuum. While there is no exact agreement between these two quantities, we find striking similarities as we vary the target state and interval size, suggesting that these quantities are closely related. 
}
\begin{document}
\maketitle

\section{Introduction}
\label{sec: Intro}

The notion of a distance between quantum states plays an important role in the subject of quantum information theory.
Several kinds of distance measures are of particular interest.
One kind is defined as a metric on the Hilbert space or on the space of mixed states whose associated geodesic distance is a measure of distinguishability of quantum states \cite{Bengtsson_Zyczkowski_2006}.
Another prominent kind of distance measure is computational complexity, or just complexity for short \cite{Papadimitriou1994}.
In practical terms, complexity measures how long it takes to run a particular algorithm on a quantum computer.
More specifically, it answers the question of how many quantum operations (taken from a limited set of operations that the quantum computer supports) are needed to prepare a given target state from a fixed reference state.

The present work is concerned with studying an example of these two kinds of distance measures in the context of the AdS/CFT correspondence.
Apart from the purely curiosity-driven question of how the distinguishability of quantum states manifests in a dual gravitational theory (in particular in the limit where quantum effects in this theory become small), the motivation for this undertaking comes from a conjecture relating computational complexity to black hole physics \cite{Susskind:2014rva,Stanford:2014jda,Brown:2015bva,Brown:2015lvg}.
In a class of quantum systems that models the physics of black holes, complexity has been observed to have certain universal features.\footnote{The models upon which the conjecture from \cite{Susskind:2014rva,Stanford:2014jda,Brown:2015bva,Brown:2015lvg} is based are finite size quantum systems whose size is proportional to the black hole entropy with time evolution modeled by a random $k$-local, all-to-all Hamiltonian, i.e.~a Hamiltonian coupling a small fixed number $k$ of randomly selected qubits at each time step.}
Loosely speaking, it shows a long period of linear growth, persisting for a time scale exponential in the black hole entropy $S_\text{BH}$, followed by saturation and fluctuations around a maximum that is exponential in $S_\text{BH}$ \cite{Susskind:2014rva,Susskind:2018pmk}.\footnote{See e.g.~\cite{Haferkamp:2021uxo,Brown:2021euk,Oszmaniec:2022srs} for more rigorous statements and proofs thereof.}
Moreover, complexity shows a characteristic response to perturbations that has been termed the ``switchback effect'' \cite{Stanford:2014jda}.
These universal features have a counterpart in AdS black hole spacetimes.
There exists a large class of geometric observables in the AdS spacetime that show the same features as complexity, i.e.~linear growth and the switchback effect \cite{Susskind:2014rva,Stanford:2014jda,Brown:2015bva,Brown:2015lvg,Couch:2016exn,Belin:2021bga,Belin:2022xmt}.
This suggests that complexity and AdS black hole physics should be connected.
More specifically, a version of the conjecture is that many geometric observable exhibiting those universal features should have a corresponding dual notion of complexity.
The class of geometric observables includes, for instance, the volume of a maximal Cauchy slice\footnote{The maximal Cauchy slice in this context is the Cauchy slice that maximizes the volume functional.} (``complexity=volume'') \cite{Susskind:2014rva,Stanford:2014jda} or the Einstein-Hilbert action on the Wheeler-de-Witt patch (``complexity=action'') \cite{Brown:2015bva,Brown:2015lvg}, but in general there are infinitely many observables that show this property \cite{Belin:2021bga,Belin:2022xmt}.
Different geometric observables (so-called holographic complexity measures) are conjectured to be dual to complexity measures with different choices of allowed set of operators or possibly different reference states according to the conjecture.\footnote{Typically, the reference state is assumed to be a spatially unentangled state \cite{Brown:2015bva,Susskind:2018pmk}. This matches the UV divergence in holographic complexity with a UV divergence from building up a large amount of entanglement inherent in any finite energy QFT state.}

Metrics on the space of states come into play in this context because they serve as building blocks for computational complexity measures.
A quantum computer supports only a discrete, finite subset of operators, so a simple direct count of the number of applied operators gives a measure of the complexity of an algorithm. In contrast, for an arbitrary quantum system, if the set of operators is infinite and possibly continuous,  a more sophisticated procedure to define complexity has to be used.
Such a procedure has been introduced by Nielsen \cite{2005quant.ph..2070N,2006Sci...311.1133N,2007quant.ph..1004D} and imported into quantum field theory in~\cite{Jefferson:2017sdb} -- see also~\cite{Chapman:2017rqy,Guo:2018kzl,Chapman:2018hou,Khan:2018rzm,Hackl:2018ptj} for further work of this approach to Nielsen complexity in quantum field theory.\footnote{Even in the discrete setting, Nielsen's complexity can serve as an upper bound on the operator counting complexity.}
In this setup, the circuit is represented by a time-ordered exponential of a time-dependent Hamiltonian instead of a discrete string of unitary operators.
A metric on the manifold of unitaries acting on the Hilbert space of the system serves as a notion of cost.
Therefore, the length of a geodesic between two points on this manifold becomes the computational complexity which, in this setting, is associated with the operator $U$ that implements the transformation from reference to target state, $\ket{\Psi_T} = U \ket{\Psi_R}$.
Unitaries that are harder to implement on the quantum computer become directions along which the metric is larger.
It is then a small leap to implement a similar measure of complexity by taking as a notion of cost a metric on the space of states instead of on the space of operators, as was done for example in \cite{Chapman:2017rqy,Yang:2017czx,Flory:2020eot,Flory:2020dja,Chagnet:2021uvi}. Thus we come to a schematic form of the definition of the complexity
\be
C(\ket{\Psi_T}, \ket{\Psi_R}) = \min_{U(t)} \int dt\,\cF(\ket{\Psi(t)}, U(t))\,, \label{eq: C-pure}
\ee
where $\cF$ is the function describing the cost of turning the state $\ket{\Psi(t)}$ into $\ket{\Psi(t+dt)}$ using the unitary operator $U(t)$.
The initial and final states $\ket{\Psi_R} = \ket{\Psi(0)}$ and $\ket{\Psi_T} = \ket{\Psi(t_f)}$ of the time evolution determine the reference and target states respectively.
In this work, we will be concerned with further generalizing this setup to mixed states.

Computational complexity for subsystems or mixed states of quantum field theories has been studied using two approaches.
The first approach extends a given definition of complexity for pure states $C_\text{pure}(\ket{\Psi_T},\ket{\Psi_R})$ to mixed states by defining $C_\text{mixed}(\rho_T,\rho_R)$ as the complexity of the associated purifications of $\rho_T,\rho_R$ minimized over all possible choices of said purifications \cite{Agon:2018zso},
\begin{equation}
    C_\text{mixed}(\rho_T,\rho_R) = \min_{\ket{\Psi_T},\ket{\Psi_R}} C_\text{pure}(\ket{\Psi_T},\ket{\Psi_R}) \quad \text{where} \quad \rho_T = \Tr_{A^c}\ket{\Psi_T},\rho_R = \Tr_{A^c}\ket{\Psi_R}.
\end{equation}
This definition is called ``purification complexity''.
Due to the vast possibilities of choosing purifications of a given mixed state, this quantity is difficult to compute even for free theories \cite{Caceres:2019pgf}.
Another definition, one that is more in tune with the idea described earlier for pure states, uses a cost function $\cF(\rho,\Phi)$, for instance, a metric on the space of mixed states as introduced above, to quantify the difficulty of applying a quantum channel on the state \cite{DiGiulio:2020hlz,DiGiulio:2021oal,DiGiulio:2021noo,Ruan:2020vze}.
Complexity is then defined as the minimal total cost for all choices of quantum circuits, that is choices of quantum channels $\Phi(t)$, that lead to the chosen target state from a given reference state,
\begin{equation}
    C_\text{mixed}(\rho_T,\rho_R) = \min_{\Phi(t)} \int dt\, \cF(\rho,\Phi(t)).
    \label{eq:complexity definition}
\end{equation}
This method is referred to as ``generalized Nielsen complexity''.
In this publication, we use the latter definition, choosing as cost function $\cF$ the Bures metric~\cite{Bures:1969}. In the context of conformal field theories, the Bures metric has been studied in \cite{Suzuki:2019xdq,Kusuki:2019hcg,Bohra:2021zyw,Erdmenger:2020vmo}, with applications for instance to entanglement wedge reconstruction in AdS/CFT \cite{Suzuki:2019xdq,Kusuki:2019hcg}.
This definition is equivalent to purification complexity for some systems if the pure state complexity definition used as input for the definition of purification complexity is based on the Fubini-Study metric as cost function \cite{Ruan:2020vze}.
Thus, our results can be thought of as applying equivalently to purification complexity or generalized Nielsen complexity.

On the gravity side, generalizations of holographic complexity measures to subsystems follow as a consequence of subregion-subregion duality~\cite{Czech:2012bh,Headrick:2013zda,Headrick:2014cta,Dong:2016eik}. The intersection of any given holographic complexity measure with the entanglement wedge of a subsystem on the AdS boundary defines a so-called holographic subregion complexity measure \cite{Alishahiha:2015rta}. The extension of the holographic complexity conjecture of \cite{Susskind:2014rva,Stanford:2014jda,Brown:2015bva,Brown:2015lvg} to this case then states that these holographic subregion complexities should be dual to complexity measures for the corresponding reduced density matrices in the dual CFT. Early works investigating properties of holographic subregion complexity include \cite{Alishahiha:2015rta,Carmi:2016wjl,Ben-Ami:2016qex}, although the subject has by now become too large to give a complete list of references. A particular interesting feature of holographic subregion complexity is that it shows a discontinuous jump at phase transition points where the location of the Ryu-Takayanagi surface changes discontinuously \cite{Ben-Ami:2016qex,Abt:2017pmf,Bhattacharya:2021jrn}.\footnote{While we will not be able to investigate these effects since we consider a single interval at zero temperature where there are no phase transitions in the entanglement entropy, they still serve as motivation for conducting our investigations as a stepping stone towards that goal.}
Qualitative comparisons to computations of subregion complexity in quantum field theories have been reported in \cite{Agon:2018zso,Caceres:2019pgf,DiGiulio:2020hlz,Camargo:2020yfv}. Some similarities between the two quantities have been found although of course a direct match cannot be expected for the non-holographic QFTs studied in these works.
In contrast, this work will focus on
any two-dimensional CFT, whether holographic or not. We will perform a
quantitative comparison to check the holographic subregion complexity conjecture.
Moreover, we will derive a bulk dual to the CFT complexity measure under consideration from first principles, using only the basic AdS/CFT dictionary.

The outline of the paper is as follows.
We start in \autoref{sec: Setup} by explaining the setup of conformal quantum circuits, as well as definitions of the Fubini-Study and Bures metrics and the associated complexity measures on spaces of pure and mixed states, respectively.
Continuing in \autoref{sec:Bures metric CFT computation}, we compute the Bures metric for a single interval in conformal quantum circuits, while \autoref{sec:Gravity dual Bures metric} describes how to map the result to the dual gravity theory.
In \autoref{sec:Holographic subregion complexity}, we calculate the holographic subregion complexity via the ``complexity=volume'' proposal and compare the result to the Bures metric complexity measure.
Finally, in \autoref{sec:Discussion}, we end with a short summary and discussion of our results. We clarify some subtleties related to the pure state limit in \autoref{app: check}, and verify that the Bures metric reduces to the Fubini-Study metric for pure states. In~\autoref{app: first law} we generalize the first law framework and leverage it to present an alternative and more general calculation to \autoref{sec:Holographic subregion complexity}.
\autoref{app:FG expansion BC} explains some details related to boundary conditions imposed in \autoref{sec:Holographic subregion complexity}, and provides useful formulas for computing holographic complexity changes in more general settings.

\section{Setup}
\label{sec: Setup}

In this section, we outline the setup of quantum circuits built out of conformal transformations that we will be using in later sections.
Moreover, we explain the definition of the Bures and Fubini-Study metrics and review previous work on complexity in quantum field theory using the Fubini-Study metric.

\subsection{Conformal quantum circuits}
We will use a framework of circuits built out of consecutive conformal transformations that was explored extensively in previous studies on the complexity of pure states \cite{Flory:2018akz,Caputa:2018kdj,Flory:2019kah,Akal:2019hxa,Erdmenger:2020sup,Flory:2020dja,Flory:2020eot,Chagnet:2021uvi,Erdmenger:2021wzc,Erdmenger:2022lov,deBoer:2023lrd}.
This means that we restrict our discussion to two-dimensional CFTs and focus on a specific class of states related by conformal transformations—i.e., arbitrary superpositions of descendants of a fixed primary state.
In this framework, a quantum circuit is defined as a sequence of states within this class. As an illustration, consider the CFT on a cylinder with Euclidean coordinates parameterized as
\be\label{eq: z zbar}
z = \phi + i t\,, \quad \bar{z} = \phi - i t\,; \quad t \in \mathbb{R}\,, \;\; \phi \in [0, 2\pi)\,.
\ee
Since conformal transformations are given by diffeomorphisms of the circle, the circuit is parametrized by a sequence of diffeomorphisms
\be
z \to w = f(t, z)\,, \quad \bar{z} \to \bar{w} = \bar{f}(t,\bar{z})\,,
\ee
where the parameter $t$ determines the position in the sequence.
The corresponding sequence of states is determined by the path ordered exponential (in Euclidean signature)
\be
\ket{\Psi(t)} = \mathcal{T} \exp\left(-\int_0^t d\tilde t H(\tilde t)\right) \ket{\Psi(0)}
\ee
where $H(t) = \int \frac{d\phi}{2\pi} T(z) \epsilon(t,z)$ and $\epsilon(t,f(t,z)) = \frac{d}{dt} f(t,z)$.

To implement the time evolution along this sequence of states, we will use the physical Hamiltonian of the CFT itself\footnote{This choice implies that we identify the circuit parameter $t$ with the physical time $(z - \bar{z}) / (2i)$. Without this identification, one needs to introduce a second Hamiltonian that governs the evolution along the sequence of states defining the circuit \cite{Erdmenger:2021wzc}.}
\be
H(t) = \int \frac{d\phi}{2\pi} \sqrt{g^{(0)}} T^t_{\phantom{t}t}.
\ee
by placing the theory on a non-trivial background $g^{(0)}_{ab}$.
As shown in \cite{Erdmenger:2021wzc}, the correct choice of background metric to implement the circuit parametrized by $f,\bar{f}$ is given by
\begin{equation}
  ds^2_{(0)} = dw d\bar{w} = \frac{\partial f}{\partial t}\frac{\partial \bar{f}}{\partial t} dt^2 + \left(\frac{\partial f}{\partial t}\frac{\partial \bar{f}}{\partial \phi} + \frac{\partial f}{\partial \phi}\frac{\partial \bar{f}}{\partial t}\right)dtd\phi + \frac{\partial f}{\partial \phi}\frac{\partial \bar{f}}{\partial \phi} d\phi^2,
  \label{eq:CFT-metric}
\end{equation}
where $w=f(t,\phi+i t)$, $\bar{w}=\bar{f}(t,\phi-i t)$.
This perspective of the circuit as the natural time evolution in a non-trivial background enables the derivation of a dual perspective on the circuit in the AdS space \cite{Erdmenger:2021wzc}. The basic idea is to use the standard AdS/CFT dictionary to derive a dual bulk geometry. As the action of a conformal transformation is generated by the non-trivial background metric \eqref{eq:CFT-metric} for the CFT which can be translated to a corresponding change in the bulk metric using the AdS/CFT dictionary, the entire quantum circuit can be described in terms of an AdS spacetime together with a choice of time-slicing \cite{Erdmenger:2021wzc}. More explicitly, changing the boundary metric corresponds to turning on a source for the boundary energy-momentum tensor. Using the Fefferman-Graham expansion, the boundary metric and the expectation value of the energy-momentum tensor determine the near-boundary behavior of the bulk metric \cite{Balasubramanian1999},
\begin{equation}
    ds^2_\text{bulk} = \frac{dr^2}{r^2} + \biggl(\frac{g_{ij}^{(0)}}{r^2} + g_{ij}^{(2)} + r^2 g_{ij}^{(4)}\biggr) dx^i dx^j
    \label{eq:bulk metric}
\end{equation}
where $g_{ij}^{(0)}$ is the boundary metric given by \eqref{eq:CFT-metric}, $g_{ij}^{(2)} = - \frac{6}{c} \ev{T_{ij}} = \frac{1}{4}\left(\frac{\partial f}{\partial x^i}\frac{\partial f}{\partial x^j} + \frac{\partial \bar f}{\partial x^i}\frac{\partial \bar f}{\partial x^j}\right)$ and $g^{(4)} = \frac{1}{4} g^{(2)} g_{(0)}^{-1} g^{(2)}$.
The time coordinate determining the circuit of interest is $t = (z-\bar{z})/(2i)$.
Note that we set the AdS radius to one in our conventions.

A peculiarity of AdS$_3$ is that the Fefferman-Graham expansion truncates, meaning that this procedure defines the entire bulk metric, not just its near-boundary expansion \cite{Skenderis2000}. Since the boundary metric \eqref{eq:CFT-metric} for a conformal quantum circuit is just a diffeomorphism of a flat metric, the bulk metric obtained by this procedure is simply pure AdS$_3$ in a particular coordinate system.\footnote{This holds true for descendants of the vacuum. The method works the same for descendants of other primaries, corresponding to conical defects of AdS$_3$ in a particular coordinate system.}
Therefore, as long as one restricts to a family of states related by conformal transformations, the choice of coordinate system (and thus the choice of time slicing) is sufficient to implement time evolution along an arbitrary sequence of states within this family.

\subsection{The Fubini-Study metric: complexity of pure states}\label{sec: Fubini-Study}
One particular choice of cost function for the definition of complexity (\ref{eq: C-pure}) that will be of interest to us is a metric on the space of pure states $\cH$ which does not assign any distance to states differing by a phase, but is otherwise homogeneous and uniform in all directions. Speaking mathematically, we consider a K\"ahler metric on the projective Hilbert space. This metric is known as the Fubini-Study metric, which in terms of pure states $\ket{\psi}\in \cH$ is written as 
\begin{equation}
    ds^2_\text{FS} = \frac{|\bracket{d\psi}{d\psi}|}{\bracket\psi\psi} - \frac{|\bracket{d\psi}{\psi}|^2}{\bracket\psi\psi^2}.
\end{equation}
This metric is the unique unitarily invariant Riemannian metric on the projective Hilbert space \cite{Bengtsson_Zyczkowski_2006}.
In geometric terms, the geodesic distance associated with the Fubini-Study metric is the angle between $\ket{\psi_1}$ and $\ket{\psi_2}$ on the Bloch sphere spanned by $\ket{\psi_1}$ on the south pole and $\ket{\chi}=(-\bracket{\psi_1}{\psi_2}\ket{\psi_1}+\ket{\psi_2})/\sqrt{1-\bracket{\psi_1}{\psi_2}-\bracket{\psi_2}{\psi_1}-|\bracket{\psi_1}{\psi_2}|^2}$ on the north pole. That the Fubini-Study metric measures distinguishability is then obvious: orthogonal states are maximally far apart in the geodesic distance while states with a large overlap are close together.
If one takes the set of allowed operations in the quantum circuit to be comprised of all operators acting on the Hilbert space of the system, this choice only leads to a trivial measure of complexity: the complexity is determined from the overlap of initial and final states and upper-bounded by $\pi/2$.

However, by restricting the set of allowed operations a striking picture emerges in the setting of 2d CFTs.
Namely, by allowing the quantum circuit to act only with conformal transformations on the reference state\footnote{This is equivalent to assigning infinite cost for every operator that is not a conformal transformation and cost given by the Fubini-Study metric for all other operators.} as explained above, many of the conjectured features of holographic complexity emerge \cite{Erdmenger:2022lov}.
The complexity for the thermofield double state dual to an eternal black hole shows linear growth for long time scales and there is an imprint of the switchback effect \cite{Erdmenger:2022lov}.
Moreover, the Fubini-Study complexity defined in this way agrees exactly with holographic complexity in the ``complexity=volume'' prescription for target states that are small conformal transformations of the vacuum up to three orders in perturbation theory~\cite{Erdmenger:2021wzc}.\footnote{To be precise, the CFT complexity agrees with the vacuum subtracted ``complexity=volume'', that is the (UV finite) difference of the volume of the maximal Cauchy slice in AdS$_3$ spacetime dual to the target state and the volume of the constant time slice of empty AdS$_3$.}

In addition, the bulk dual to the conformal quantum circuits outlined above allows for translating the Fubini-Study complexity measure to a simple geometric quantity in this dual spacetime.
In that way, the AdS dual to the Fubini-Study complexity was derived from first principles in \cite{Erdmenger:2022lov}.
We will similarly provide a first-principles derivation of an AdS dual to the Bures metric in \autoref{sec:Gravity dual Bures metric}.

\subsection{The Bures metric: complexity of mixed states}

We now turn our attention to studying subsystems of a larger system and distance measures for the associated reduced density matrices. We do this by the ``generalized Nielsen complexity'' framework described in the introduction, which consists of defining a cost function ${\cal F}(\rho,\Phi)$ in the space of mixed states, and minimizing the cost of the quantum chanel $\Phi(t)$ needed to reach a target state $\rho_T$ from a reference state $\rho_R$ -- see eq.~\eqref{eq:complexity definition}.

Metrics on the space of mixed states are well-studied in quantum information theory (see \cite{Bengtsson_Zyczkowski_2006} for an extensive summary).
As mentioned in~\autoref{sec: Fubini-Study}, there is a unique metric on the projective Hilbert space that does not distinguish any particular direction -- the Fubini-Study metric. 
However, there is no unique generalization of this for mixed states \cite{Bengtsson_Zyczkowski_2006}.
To quantify the distinguishability of two density matrices $\rho$ and $\sigma$, it is useful to introduce the notion of a monotone distance $D$ in the space of density matrices characterized by the property that $D(\Phi(\rho),\Phi(\sigma)) \leq D(\rho,\phi)$.
Here, the quantum channel $\Phi(\rho)$, also called stochastic map, is the most general operation one can perform on a quantum state, i.e.~a combination of unitary evolution and measurement.\footnote{To be specific, to implement a quantum channel one introduces ancillary degrees of freedom, performs a unitary evolution on the total system (original degrees of freedom plus ancillae) and traces out the ancillary system again such that $\rho \to \sum_i A_i \rho A_i^\dagger$ where $\sum_i A_i^\dagger A_i = 1$.}
This condition ensures that the distance between the two quantum states $\rho$ and $\sigma$ does not increase under any physical operation and thus provides a measure of distinguishability.
There are infinitely many Riemannian metrics whose geodesic distance is monotone.
A well-known example we will study in this publication is the Bures metric \cite{Bures:1969}.\footnote{To give some intuition on this metric, we note that for a two-level system, where the Bloch ball gives the space of mixed states, it takes the form $ds^2_B = \frac{1}{4}\left[\frac{dr^2}{1-r^2}+r^2d\Omega^2\right]$ \cite{Bengtsson_Zyczkowski_2006}.
It reduces to the Fubini-Study metric for pure states ($r=1$).
More general metrics for two-level systems are determined by adjusting the prefactor of the angular part $d\Omega^2$.}
We now proceed to introduce this metric, which is a crucial tool for quantifying distances in the space of density matrices and which will be essential for our study of complexity in mixed states.

Consider two density matrices, $\rho$ and $\sigma$, which represent arbitrary mixed states in a QFT. The Bures metric is defined in terms of the fidelity functional, which measures the overlap between $\rho$ and $\sigma$
\be
F(\rho,\sigma) = \left[\Tr\sqrt{\sqrt{\rho}\sigma\sqrt{\rho}}\right]^2\,. \label{eq: fidelity}
\ee
Using the fidelity, the Bures distance is given by
\be
D(\rho, \sigma) = 2 (1-F(\rho, \sigma))\,.
\ee
In this work, we are particularly interested in the density matrices of a subsystem $A$, which resides within a subregion of a Cauchy slice. Assuming that the full QFT can be purified, the total state of the system is described by a pure state $|\Psi\rangle$, with the corresponding density matrix
\be
\rho_{\text{tot}}(t) = | \Psi(t) \rangle \langle \Psi(t) |\,.
\ee
The density matrix of the subsystem $A$ is then obtained by tracing out the degrees of freedom of the complement, $\overline{A}$, as follows:
\be
\rho(t) = \Tr_{\overline{A}} \rho_{\text{tot}}(t)\,.
\ee
Next, we consider the time evolution of the subsystem density matrix $\rho$ under the Hamiltonian $H$ of the CFT, which evolves as
\be
\rho(t) = \Tr_{\overline{A}} \left[ e^{i H t} \rho_{\text{tot}}(0) e^{-i H t}\right]\,.
\ee
Our goal is to compute the Bures metric along the time evolution direction in the space of density matrices
\begin{equation}
	ds^2_B = \cF_B(t) dt^2 = -\partial_{\delta t}^2 \left.F(\rho(t),\rho(t+\delta t))\right|_{\delta t=0} dt^2 \,.
	\label{eq:Bures-metric}
\end{equation}
To calculate the derivative of the fidelity, we will use a slightly more general functional known as the quantum Renyi relative entropy \cite{Lashkari:2014yva}
\begin{equation}
	S_\alpha(\sigma\, ||\, \rho) = \frac{1}{\alpha-1} \log \Tr_A \left[ \left(\rho^{\frac{1-\alpha}{2\alpha}}\sigma\rho^{\frac{1-\alpha}{2\alpha}} \right)^\alpha \right]\,.
\end{equation}
For $\alpha = 1/2$, this reduces to the fidelity, and the Bures metric is then given by
\begin{equation}\label{eq: bures metric}
	\cF_B(t) \equiv- \left.\partial_{\delta t}^2 S_\alpha(\rho(t+\delta t)\, ||\, \rho(t))\right|_{\delta t=0,\alpha=1/2}\,.
\end{equation}
This will be the starting point for the CFT computation of the Bures metric, which we carry in the following section.

\section{CFT computation of the Bures metric}
\label{sec:Bures metric CFT computation}
We compute the Bures metric \eqref{eq: bures metric} for mixed states in a $2$d CFT using the replica trick. In this section, we outline how the Bures metric can be evaluated using this method, starting from the expansion of the time-evolved density matrix $\rho(t)$ and calculating the fidelity functional between $\rho(t)$ and $\rho(0)$ through the replica trick. The process involves first computing the fidelity for integer replica parameters and then analytically continuing the result to non-integer values.

We begin by expanding the time-evolved density matrix $\rho(t)$, which describes the time evolution of the subsystem state
\begin{equation}
	\begin{aligned}
		\rho(t+\delta t) &= \Tr_{\bar A}\left[e^{i H \delta t} \rho_\text{tot}(t) e^{-i H \delta t}\right] = \rho(t) + \delta t\ \rho_1(t) + \delta t^2 \rho_2(t) + ...,\\
  \rho(t) &=  \Tr_{\bar A} \rho_\text{tot}(t),\\
		\rho_1(t) &= i \Tr_{\bar A}[H, \rho_\text{tot}(t)],\\
		\rho_2(t) &= \frac{1}{2}\Tr_{\bar A}[-H^2 \rho_\text{tot}(t) + 2H \rho_\text{tot}(t) H - \rho_\text{tot}(t) H^2].
	\end{aligned}
	\label{eq:definition-rho-rho1-rho2}
\end{equation}
Here, $\rho$ is the unperturbed density matrix, and $\rho_1$ and $\rho_2$ are the first and second-order corrections due to time evolution.

We aim to evaluate the Bures metric \eqref{eq: bures metric} which is defined as the second derivative of the fidelity between $\rho(t+\delta t)$ and $\rho(t)$. The fidelity between the two density matrices can be conveniently computed using the replica trick. Specifically, we express the fidelity in terms of the replica parameters $n,m$ as
  \begin{equation}
	F(\rho(t+\delta t),\rho(t)) = \lim_{n,m \to 1/2} \left[\Tr[(\rho(t)^m \rho(t+\delta t) \rho(t)^m)^n]\right]^2\,.
	\label{eq:fidelity-replica-trick}
\end{equation}

To proceed, we compute 
$\partial_{\delta t}^2\left.\Tr[(\rho(t)^m \rho(t+\delta t) \rho(t)^m)^n]\right|_{\delta t=0}$ 
for integer values of $n$ and $m$, and then analytically continue the result to $n,m\to1/2$, as required by the replica trick. We begin by defining
\begin{equation}
  G_{mn}(t,\delta t) \equiv 
  \Tr_A[(\rho^m (\rho + \delta t\rho_1 + \delta t^2\rho_2) \rho^m)^n]
  \,,
\end{equation}
for integer $m,n$, where $\rho,\rho_1$ and $\rho_2$ are defined in \eqref{eq:definition-rho-rho1-rho2} and are evaluated at time $t$. In what follows, all density matrices are evaluated at time $t$ unless otherwise specified. The first derivative of $G_{nm}$ vanishes as $n,m\to1/2$
\begin{equation}
  \begin{aligned}
    \lim_{n,m \to 1/2}\left.\partial_{\delta t} G_{nm}
    \right|_{\delta t=0} 
    &= \lim_{n,m \to 1/2} \sum_{k=1}^n \Tr_A[\rho^{(k-1)(2m+1)}\rho^m\rho_1\rho^m\rho^{(n-k)(2m+1)}]\,, \\
    &=  \lim_{n,m \to 1/2} n\Tr_A[\rho^{n(2m+1)-1}\rho_1]\,,\\
    &= \frac{1}{2}\Tr_A[\rho_1]\,,\\
    &= 0 \,,
  \end{aligned}
\end{equation}
which is expected because the fidelity in~\eqref{eq:fidelity-replica-trick} is maximized at $\delta t=0$.

The second derivative is given by
\begin{equation}\label{eq: Bures derivative}
  \left.\partial^2_{\delta t} G_{nm}\right|_{\delta t=0} = 2n \Tr_A[\rho^{n(2m+1)-1}\rho_2] + n \sum_{k=1}^{n-1} \Tr_A[\rho^{k(2m+1)-1}\rho_1\rho^{(n-k)(2m+1)-1}\rho_1].
\end{equation}
After analytically continuing to $n,m \to 1/2$, the first term involving $\Tr_A[\rho^{n(2m+1)-1}\rho_2]$ drops out since $\Tr_A[\rho_2] = 0$ (see \autoref{app: check} for more details) leaving us with
\be\label{eq: G_nm}
\lim_{n,m \to 1/2}\left.\partial^2_{\delta t} G_{nm}\right|_{\delta t=0}  =\lim_{n,m \to 1/2} n \sum_{k=1}^{n-1} \Tr_A[\rho^{k(2m+1)-1}\rho_1\rho^{(n-k)(2m+1)-1}\rho_1]\,.
\ee
At this stage, the derivation has been completely general. To evaluate this trace explicitly, we will restrict ourselves to $2$d CFTs in what follows.

\subsection{Computing the replicated Bures metric}

\label{sec:CFT-computation}
Let us now explain how to compute traces of the form
\begin{equation}
\Tr_A[\rho^{n_1}\rho_1\rho^{n_2}\rho_1] \quad \text{for} \quad n_{1,2} \in \ZZ_+.
\end{equation}
We will consider this computation for the case of the vacuum state $| \Psi \rangle = | 0 \rangle$ of a 2$d$ CFT, so that $\rho_{\text{tot}} = | 0 \rangle \langle 0 |$. First, let us set $n_1=n_2=0$. Using the form $H = \int_0^{2\pi} dw \sqrt{\det g(w)} T^t_t(w)$ of the Hamiltonian gives
\begin{equation}
  \begin{aligned}
    \Tr_A[\rho_1^2] &= -\Tr_A[\Tr_{\bar A}[H,\ket{0}\bra{0}]^2]\\
    &= - \int_0^{2\pi} dw_1 dw_2 \sqrt{\det g(w_1)}\sqrt{\det g(w_2)} \Tr_A[\Tr_{\bar A}[T^t_t(w_1),\ket{0}\bra{0}]\Tr_{\bar A}[T^t_t(w_2),\ket{0}\bra{0}]].
  \end{aligned}
    \label{eq:derivation-replica-correlator-1}
\end{equation}
To compute the trace over $A$, we insert a complete set of states as follows
\begin{equation}
    \begin{aligned}
    &\Tr_A[\Tr_{\bar A}[T^t_t(w_1),\ket{0}\bra{0}]\Tr_{\bar A}[T^t_t(w_2),\ket{0}\bra{0}]]\\
    & = \int dX_1 dX_2 \bra{X_1}\Tr_{\bar A}[T^t_t(w_1),\ket{0}\bra{0}]\ket{X_2}\bra{X_2}\Tr_{\bar A}[T^t_t(w_2),\ket{0}\bra{0}]\ket{X_1}.
    \end{aligned}
    \label{eq:derivation-replica-correlator-2}
\end{equation}
The correlators $\bra{X_1}\Tr_{\bar A}[T^t_t(w_1),\ket{0}\bra{0}]\ket{X_2}$ can be computed in terms of an Euclidean path integral
\begin{equation}
    \begin{aligned}
    \bra{X_1}\Tr_{\bar A}[T^t_t(w_1),\ket{0}\bra{0}]\ket{X_2} = \int &[\cD X] e^{-S[X]} (T^t_t(t=0^-,w_1) - T^t_t(t=0^+,w_1))\\
    &\times\prod_{w \in A}\delta(X(t=0^+,w)-X_2(w))\delta(X(t=0^-,w)-X_1(w))\,,
    \end{aligned}
    \label{eq:derivation-replica-correlator-3}
\end{equation}
where $X$ stands for the entire set of fields of the CFT, and $S[X]$ is the Euclidean action of the CFT. Because the correlator is time ordered, the insertions of the energy-momentum tensor approaching $t=0$ from below or above implement the ordering of the terms in the commutator $[T^t_t(w_1),\ket{0}\bra{0}]$.
In \eqref{eq:derivation-replica-correlator-2}, we have two copies of this path integral.
The fields in these copies are sewn together by the cyclicity of the trace.

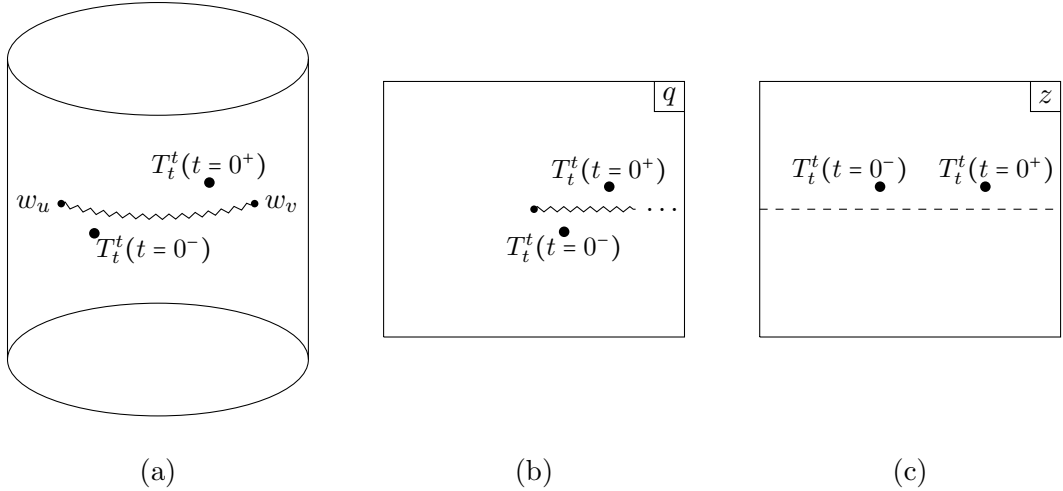
\begin{figure}
  \centering
  \begin{tikzpicture}

    \begin{scope}[shift={(-5,0)}]
      \draw (0,-2) ellipse (2 and 0.75);
      \draw (0,2) ellipse (2 and 0.75);
      \draw (2,-2) -- (2,2);
      \draw (-2,-2) -- (-2,2);
      \draw[decorate,decoration={zigzag,amplitude=1,segment length=5}] ({{-2*sin(40)}},{{0.75*cos(40)-0.5}}) arc(230:310:2 and 0.75) ({{2*sin(40)}},{{0.75*cos(40)-0.5}});
      \fill ({{-2*sin(40)}},{{0.75*cos(40)-0.5}}) circle(0.05);
      \draw ({{-2*sin(40)}},{{0.75*cos(40)-0.5}}) node[left] {$w_u$};
      \fill ({{2*sin(40)}},{{0.75*cos(40)-0.5}}) circle(0.05);
      \draw ({{2*sin(40)}},{{0.75*cos(40)-0.5}}) node[right] {$w_v$};
      \fill ({{-2*sin(25)}},{{0.75*cos(25)-1}}) circle(0.07);
      \draw ({{-2*sin(25)}},{{0.75*cos(25)-1}}) node[below right,inner sep=0.5] {\small $T^t_t(t=0^-)$};
      \fill ({{2*sin(20)}},{{0.75*cos(20)-0.35}}) circle(0.07);
      \draw ({{2*sin(20)}},{{0.75*cos(20)-0.35}}) node[above,inner sep=1] {\small $T^t_t(t=0^+)$};
      \node at (0,-3.5) {(a)};
    \end{scope}

    \begin{scope}[shift={(0,0)}]
      \draw (-2,-1.7) -- (2,-1.7) -- (2,1.7) -- (-2,1.7) -- (-2,-1.7);
      \draw (1.8,1.5) node {$q$};
      \draw (1.6,1.7) -- (1.6,1.3) -- (2,1.3);
      \fill (0,0) circle(0.05);
      \draw[decorate,decoration={zigzag,amplitude=1,segment length=5}] (0,0) -- (1.35,0);
      \draw (1.35,0) node[right] {$\dots$};
      \fill (0.4,-0.3) circle(0.07);
      \draw (0.4,-0.3) node[below,inner sep=0.5] {\small $T^t_t(t=0^-)$};
      \fill (1.0,0.3) circle(0.07);
      \draw (1.0,0.3) node[above,inner sep=0.5] {\small $T^t_t(t=0^+)$};
      \node at (0,-3.5) {(b)};
    \end{scope}

    \begin{scope}[shift={(5,0)}] 
      \draw (-2,-1.7) -- (2,-1.7) -- (2,1.7) -- (-2,1.7) -- (-2,-1.7);
      \draw (1.8,1.5) node {$z$};
      \draw (1.6,1.7) -- (1.6,1.3) -- (2,1.3);
      \draw[dashed] (-2,0) -- (2,0);
      \fill (-0.4,0.3) circle(0.07);
      \draw (-0.4,0.3) node[above,inner sep=1,shift={(-0.4,0)}] {\small $T^t_t(t=0^-)$};
      \fill (1,0.3) circle(0.07);
      \draw (1,0.3) node[above,inner sep=1,shift={(0.15,0)}] {\small $T^t_t(t=0^+)$};
      \node at (0,-3.5) {(c)};
    \end{scope}

  \end{tikzpicture}

\caption{
    The three-step process of mapping a two-sheeted Riemann surface to the complex plane when computing $\Tr_A[\rho_1^2]$. \textbf{(a)}: The initial geometry, a two-sheeted Riemann surface with operator insertions $T^t_t(t=0^+)$ and $T^t_t(t=0^-)$ on the first sheet near the branch points.  \textbf{(b)}: The first conformal map, $\frac{\sin\left(\frac{w-w_u}{2}\right)}{\sin\left(\frac{w-w_v}{2}\right)} =q $, maps the surface to a two-sheeted covering of the complex plane where the branch cut runs from $0$ to $\infty$. \textbf{(c)}: The second map, $\sqrt{q} =z$, transforms the geometry to a single complex plane where the two Riemann sheets are represented by the upper and lower half-planes. The operator insertions originally located above and below the branch cut in (b) are mapped to insertions just above the dashed line in (c) separating the upper and lower half planes.}
  \label{fig:replica-trick}
\end{figure}

Therefore, we find that $\Tr_A[\rho_1^2]$ is equal to a two-point correlator
\begin{equation} \label{eq:two-point-correlator-Riemann-surface}
\begin{aligned}
     \Tr_A[\rho_1^2]= - \int_{w_u}^{w_v} &dw_1 dw_2 \sqrt{\det g(w_1)}\sqrt{\det g(w_2)}\\
     &\ev{(T^t_t(t=0^-,w_1) - T^t_t(t=0^+,w_1))(T^t_t(t=0^-,w_2) - T^t_t(t=0^+,w_2))}\,,
\end{aligned}
\end{equation}
on a two-sheeted Riemann surface with a branch cut along the interval $A=[w_u,w_v]$ (see \autoref{fig:replica-trick} (a)) where the two insertions at $w=w_1$ and $w=w_2$ lie on the two different sheets, respectively.
One can restrict to $w_{1,2} \in [w_u,w_v]$ because the difference $T^t_t(t=0^-,w_1) - T^t_t(t=0^+,w_1)$ between the energy-momentum tensor above and below the Euclidean time $t=0$ is non-vanishing only along the branch cut.

The two-sheeted Riemann surface is mapped to the complex plane using the uniformization map
\begin{equation}
  z = \sqrt{\frac{\sin((w-w_u)/2)}{\sin((w-w_v)/2)}}.
\end{equation}
This is illustrated in \autoref{fig:replica-trick}. 
The energy-momentum tensor transforms as usual under this map,
\begin{equation}
  T(w) = \left(\frac{\partial z}{\partial w}\right)^2 T(z) + \frac{c}{12}\{z,w\}.
\end{equation}
and the correlator on the complex plane takes the well-known form
\begin{equation}
  \ev{T(z_1) T(z_2)} = \frac{c/2}{(z_1-z_2)^4}.
\end{equation}

The calculation for general $n_{1,2}$ works analogously.
In this case, the Riemann surface has $n_1+n_2+2$ sheets.
Without loss of generality, one energy-momentum tensor insertion can be taken to lie on the first sheet while the other lies on the $(2+n_2)$'th sheet. The generalization of eq.~\eqref{eq:two-point-correlator-Riemann-surface} is
\begin{equation} \label{eq:two-point-correlator-Riemann-surface gen}
\begin{aligned}
     \Tr_A[\rho^{n_1}\rho_1\rho^{n_2}\rho_1]= - \int_{w_u}^{w_v} &dw_1 dw_2 \sqrt{\det g(w_1)}\sqrt{\det g(w_2)}\\
     &\ev{(T^t_t(t=0^{-},w_1) - T^t_t(t=0^{+},w_1))(T^t_t(t=0^{-},w_2) - T^t_t(t=0^{+},w_2))}\,,
\end{aligned}
\end{equation}
where the two insertions on the different sheets are labeled by $w=w_1$ and $w=w_2$, respectively.
The uniformization map is given by
\begin{equation}
  z = \left(\frac{\sin((w-w_u)/2)}{\sin((w-w_v)/2)}\right)^{1/(n_1+n_2+2)}.
\end{equation}
Therefore, the two-point correlator on the Riemann surface is given by
\begin{equation}
  \begin{aligned}
    \ev{T(w_1)T(w_2)} = ~&\frac{c}{32(n_1+n_2+2)^4}\frac{\sin^4\bigl(\frac{w_u-w_v}{2}\bigr)}{\sin^2\bigl(\frac{w_1-w_u}{2}\bigr)\sin^2\bigl(\frac{w_1-w_v}{2}\bigr)\sin^2\bigl(\frac{w_2-w_u}{2}\bigr)\sin^2\bigl(\frac{w_2-w_v}{2}\bigr)}\\
                         &\qquad\times\left[e^{-i\alpha}x^{\frac{1}{2(2+n_1+n_2)}} -e^{i\alpha}x^{-\frac{1}{2(2+n_1+n_2)}}\right]^{-4}\\
    &+\left(\frac{c}{12}\right)^2\{z_1,w_1\}\{z_2,w_2\}\,,
  \end{aligned}
\end{equation}
where we have used that the insertions are on different sheets.
The conformal cross-ratio $x$ is defined by
\begin{equation}\label{eq: x equation}
x=\frac{\sin((w_1-w_u)/2)\sin((w_2-w_v)/2)}{\sin((w_1-w_v)/2)\sin((w_2-w_u)/2)}\,.
\end{equation}
The phase $\alpha =  \frac{N \pi}{n_1+n_2+2}$ is there because of the separation between the insertion points in the $z$ plane. The integer $N$ depends on whether the stress tensor insertions are at $t = 0^+$ or $t=0^-$.  In our conventions, 
\begin{equation}
    N = \begin{cases}
    n_2 \quad \quad \ {\rm for} \quad  t_1=0^-\,, t_2=0^+\,,\\
    n_2+1 \quad {\rm for} \quad t_1=0^+\,, t_2=0^+\,,\ {\rm and}\quad t_1=0^-\,, t_2=0^-\,,
    \\
    n_2+2 \quad {\rm for} \quad t_1=0^+\,, t_2=0^-\,.
    \end{cases}
\end{equation}
To obtain $\Tr_A[\rho^{n_1}\rho_1\rho^{n_2}\rho_1]$, we insert this result into \eqref{eq:two-point-correlator-Riemann-surface gen} and use the form of the CFT metric \cite{Erdmenger:2022lov} given by \eqref{eq:CFT-metric}.

One can use the results obtained in this section to further simplify equation \eqref{eq: G_nm}. Specifically, we find that for the conformally transformed vacuum state 
\begin{equation}\label{eq:to continue}
	\begin{aligned}
		\lim_{n,m \to 1/2}\left.\partial^2_{\delta t} G_{nm}\right|_{\delta t=0} &=\lim_{n,m \to 1/2}n\sum_{k=1}^{n-1} \Tr_A[\rho^{k(2m+1)-1}\rho_1\rho^{(n-k)(2m+1)-1}\rho_1]\\
		&=-\lim_{n,m \to 1/2}n \int_{w_u}^{w_v} dw_1 dw_2\, \partial_{t_1}f_1 \partial_{t_2}f_2 \frac{c}{16n^4(2m+1)^4}\\
  &\qquad\times\frac{\sin^4\bigl(\frac{w_u-w_v}{2}\bigr)}{\sin^2\bigl(\frac{w_1-w_u}{2}\bigr)\sin^2\bigl(\frac{w_1-w_v}{2}\bigr)\sin^2\bigl(\frac{w_2-w_u}{2}\bigr)\sin^2\bigl(\frac{w_2-w_v}{2}\bigr)}\\
		&\qquad\times\sum_{k=1}^{n-1}\biggl[
		\begin{aligned}[t]
			&\left(x^{\frac{1}{2n(2m+1)}}e^{i\pi\frac{k}{n}}-x^{-\frac{1}{2n(2m+1)}}e^{-i\pi\frac{k}{n}}\right)^{-4}\\
			&-\left(x^{\frac{1}{2n(2m+1)}}e^{i\pi\left(\frac{k}{n}+\frac{1}{n(2m+1)}\right)}-x^{-\frac{1}{2n(2m+1)}}e^{-i\pi\left(\frac{k}{n}+\frac{1}{n(2m+1)}\right)}\right)^{-4}\biggr] \\
   &\qquad+ \text{antihol.~part}
		\end{aligned}
	\end{aligned}
\end{equation}
where we introduced the notation $f_{1,2}=f(t,\phi_{1,2}+i t)$ while $w_u=f(t,u+i t)$ and $w_v=f(t,v+i t)$ denote the endpoints of the interval $\phi \in [u,v]$ in $w$ coordinates. Note that the part involving the Schwarzian cancels in~\eqref{eq:two-point-correlator-Riemann-surface}. The functions $f$ and $\bar{f}$ parametrize the conformal transformation applied at each time $t$ and therefore depend explicitly on $t$ as well as on the respective (anti-)holomorphic coordinate $\phi \pm i t$.

\subsection{Analytic continuation}

To simplify the expression obtained for the Bures metric in \eqref{eq:to continue}, we notice that the terms in the sum are both of the following form

\begin{equation}
    \left(y^{\frac{1}{2n(2m+1)}}e^{\frac{k \pi i}{n}}-y^{-\frac{1}{2n(2m+1)}}e^{-\frac{k \pi i}{n}}\right)^{-4}\,.
\end{equation}
Specifically, the two terms in \eqref{eq:to continue} can be identified through the redefinitions $y= x $ and $y=x e^{2\pi i}$ respectively. This observation allows us to further simplify the sum in \eqref{eq:to continue} by focusing on the computation of the following key expression
\begin{equation}
    S(y)=\sum_{k=1}^{n-1}\left(y^{\frac{1}{2n(2m+1)}}e^{\frac{k \pi i}{n}}-y^{-\frac{1}{2n(2m+1)}}e^{-\frac{k \pi i}{n}}\right)^{-4}\,.
\end{equation}

Since we are interested in the behavior as $n,m\to1/2$, we begin by considering the sum $S(y)$ for $m=1/2$. This leads to the following expression
\begin{equation}
    \lim_{m \to 1/2}S(y)=\sum_{k=1}^{n-1}\left(y^{\frac{1}{4n}}e^{\frac{k \pi i}{n}}-y^{-\frac{1}{4n}}e^{-\frac{k \pi i}{n}}\right)^{-4}\,.
\end{equation}
For $|y|<1$, this sum can be computed as follows
\begin{equation}
  \begin{aligned}
      \lim_{m \to 1/2}S(y)
    &=\sum_{k=1}^{n-1}
    y^{\frac{1}{n}}e^{\frac{4i\pi k}{n}} \left(1-y^{-\frac{1}{2n}}e^{\frac{-2i\pi k}{n}}\right)^{-4}\\
    &=\sum_{k=1}^{n-1}
    y^{\frac{1}{n}}e^{\frac{4i\pi k}{n}}\sum_{q=0}^{\infty} \frac{(q+3)(q+2)(q+1)}{6}y^{\frac{q}{2n}}e^{\frac{2i\pi q k}{n}}\\
    &=
    y^{\frac{1}{n}}\sum_{q=0}^{\infty} \frac{(q+3)(q+2)(q+1)}{6}y^{\frac{q}{2n}}\sum_{k=1}^{n-1}e^{\frac{2i\pi (q+2) k}{n}}\,.
  \end{aligned}
\end{equation}
In the second line, we have used that for $|x|<1$,
\begin{equation}
    \sum_{q=0}^{\infty} \frac{(q+3)(q+2)(q+1)}{6}x^{q}=(x-1)^{-4}\,.
\end{equation}
In the last line, we changed the order of summation. 

Furthermore, we know that $\sum_{k=1}^{n-1}e^{\frac{2ik\pi j}{n}}$ gives $n-1$ if $j=0 \Mod n$, and $-1$ otherwise. This implies
\begin{equation}
  \begin{aligned}
     \lim_{m \to 1/2}S(y)
       &=
       n y^{\frac{1}{n}}\sum_{q=0}^{\infty} \frac{(q n+1)(qn)(qn-1)}{6}y^{\frac{qn-2}{2n}}\\
       &\quad-y^{\frac{1}{n}}\sum_{q=0}^{\infty} \frac{(q+3)(q+2)(q+1)}{6}y^{\frac{q}{2n}}\,.
  \end{aligned}
\end{equation}
In the first sum, we have changed to summation index $q\to qn-2$ to simplify the expression.

Assuming that $|y|<1$, this sum can be performed and we find
\begin{equation}
  \begin{aligned}
      \lim_{m \to 1/2}S(y)
       &=
      \frac{n^2 \sqrt{y} \left(-\left(\sqrt{y}-1\right)^2+n^2 \left(y+4 \sqrt{y}+1\right)\right)}{6 \left(\sqrt{y}-1\right)^4}-\frac{y^{1/n}}{\left(-1+ y^{\frac{1}{2 n}}\right)^4}\,.
  \end{aligned}
\end{equation}
The same expression can be found for $|y|>1$ by changing $y\to y^{-1}$.

Finally, we can take $n=\frac{1}{2}$ and find that
\begin{equation}\label{eq: final sum}
  \begin{aligned}
      \lim_{n,m \to 1/2}S(y)=\sum_{k=1}^{n-1}\left(y^{\frac{1}{4n}}e^{\frac{k \pi i}{n}}-y^{-\frac{1}{4n}}e^{-\frac{k \pi i}{n}}\right)^{-4}
      &=-\frac{y^{3/2}+4 y+\sqrt{y}}{32 \left(1+\sqrt{y}\right)^4}\,.
  \end{aligned}
\end{equation}
Now we can go back to the sum in~\eqref{eq:to continue} and use the result obtained in \eqref{eq: final sum} for $y\to x$ and $y\to x e^{2\pi i}$, to obtain that
\begin{equation}
        \sum_{k=1}^{n-1}\biggl[
		\begin{aligned}[t]
			&\left(x^{\frac{1}{2n(2m+1)}}e^{i\pi\frac{k}{n}}-x^{-\frac{1}{2n(2m+1)}}e^{-i\pi\frac{k}{n}}\right)^{-4}\\
			&-\left(x^{\frac{1}{2n(2m+1)}}e^{i\pi\left(\frac{k}{n}+\frac{1}{n(2m+1)}\right)}-x^{-\frac{1}{2n(2m+1)}}e^{-i\pi\left(\frac{k}{n}+\frac{1}{n(2m+1)}\right)}\right)^{-4}\biggr] \\
   = & -\frac{x^{3/2}+4 x+\sqrt{x}}{32 \left(\sqrt{x}+1\right)^4}-\frac{x^{3/2}-4 x+\sqrt{x}}{32 \left(\sqrt{x}-1\right)^4} \\
   = &- \frac{\sqrt{x} \left(x^3-9 x^2-9 x+1\right)}{16 (x-1)^4}
    \end{aligned}
\end{equation}
Therefore the Bures metric~\eqref{eq:to continue} is
\begin{equation}\label{eq: Bures metric}
  \begin{aligned}
   \cF_B  &= \int_{w_u}^{w_v} dw_1 dw_2\, \partial_{t_1}f_1\partial_{t_2}f_2 \frac{c}{32}\frac{\sin^4\bigl(\frac{w_u-w_v}{2}\bigr)}{\sin^2\bigl(\frac{w_1-w_u}{2}\bigr)\sin^2\bigl(\frac{w_1-w_v}{2}\bigr)\sin^2\bigl(\frac{w_2-w_u}{2}\bigr)\sin^2\bigl(\frac{w_2-w_v}{2}\bigr)}\\ 
   &\times\biggl[
      \begin{aligned}[t]
        &-\frac{\sqrt{x} (1 - 9 x - 9 x^2 + x^3)}{16(x-1)^4}\biggr] + \text{antihol.~part}
        \end{aligned} \\
    &= \int_{u}^{v} d\phi_1 d\phi_2\, \partial_{\phi_1}f_1\partial_{t_1}f_1\partial_{\phi_2}f_2\partial_{t_2}f_2 \frac{c}{32}\frac{\sin^4\bigl(\frac{w_u-w_v}{2}\bigr)}{\sin^2\bigl(\frac{w_1-w_u}{2}\bigr)\sin^2\bigl(\frac{w_1-w_v}{2}\bigr)\sin^2\bigl(\frac{w_2-w_u}{2}\bigr)\sin^2\bigl(\frac{w_2-w_v}{2}\bigr)}\\
    &\times\biggl[
      \begin{aligned}[t]
        &-\frac{\sqrt{x} (1 - 9 x - 9 x^2 + x^3)}{16(x-1)^4}\biggr] + \text{antihol.~part}\,,
      \end{aligned}
  \end{aligned}
\end{equation}
where recall that we introduced the notation $f_{1,2}=f(t,\phi_{1,2}+i t)$, $w_u=f(t,u+it)$ and $w_v=f(t,v+it)$.
This is one of the main results of this work. 

Let us now check if the expression correctly reduces to the Fubini-Study distance in the limit that the interval $A$ covers the entire boundary. For pure states $\rho = \ket\Psi\bra\Psi$ on the entire space, we have $v=u+ 2\pi$. In this case, we find that
\begin{equation}\label{eq: reduce to FS}
  \begin{aligned}
    \lim_{v \to u+ 2\pi}\cF_B 
      &=\int_0^{2\pi} d\phi_1 d\phi_2\, \partial_{\phi_1}f_1\partial_{t_1}f_1\partial_{\phi_2}f_2\partial_{t_2}f_2 \frac{c}{32}\frac{1}{\sin^4\bigl(\frac{w_1-w_2}{2}\bigr)}+ \text{antihol.~part}\,,\\
      &=\cF_\text{FS}\,.
  \end{aligned}
\end{equation}
Hence, as expected the result is in agreement with the Fubini-Study metric \cite{Erdmenger:2022lov}. We refer the reader to \autoref{app: check} for more details on the recovery of the Fubini-Study metric from the Bures metric.

\subsection{Bures metric in Fourier basis}

To proceed with the integration in the expression obtained for the Bures metric~\eqref{eq: Bures metric}, we use that $\partial_t f$ is periodic in $w$ and thus can be expanded in a Fourier series as follows
\begin{equation}\label{eq:f1 fourier}
    \partial_t f(t,\phi(w)+i t) = \sum_{n=-\infty}^\infty f^n e^{i n w}\,.
\end{equation}
Therefore, the Bures metric is given by
\begin{equation}
    \cF_B =-\frac{c}{512}\sum_{n,m}f^n f^m S_{nm}
\end{equation}
where the matrix elements $S_{nm}$ encode the contributions from the various Fourier components and are defined as 
\begin{equation}\label{eq: Smn}
\begin{aligned}
    S_{nm}  = & \int_{w_u}^{w_v} dw_1 dw_2 \frac{ e^{i n w_1 + i m w_2}\sin^4\bigl(\frac{w_u-w_v}{2}\bigr)}{\sin^2\bigl(\frac{w_1-w_u}{2}\bigr)\sin^2\bigl(\frac{w_1-w_v}{2}\bigr)\sin^2\bigl(\frac{w_2-w_u}{2}\bigr)\sin^2\bigl(\frac{w_2-w_v}{2}\bigr)}
   \frac{\sqrt{x} (1 - 9 x - 9 x^2 + x^3)}{(x-1)^4} \\ 
   & + {\rm antihol.\ part}\,.
\end{aligned}
\end{equation}
Importantly, the conformal transformation~\eqref{eq:f1 fourier} is real, so the coefficients $f^n$ are not all independent, and in particular
\begin{equation}\label{eq: reality}
    f^{-n}=f^{n*}\,.
\end{equation}

\begin{figure}
\centering
\begin{tikzpicture}

    \draw[black] (-3.5, -3.5) rectangle (3.5, 3.5);

    \draw[black] (3, 3) rectangle (3.5, 3.5);
    \node at (3.25, 3.25) {\small $W_1$};

    \draw[black] (-3.5, 0) -- (3.5, 0);  
    \draw[black] (0, -3.5) -- (0, 3.5);  

    \draw[black, thick] (0,0) circle (2);

    \draw[blue, thick] (0,0) circle (3);

    \node[blue] at (2.4,2.4) {\large $C_2$};

    \coordinate (u) at (-1.5, 1.34); 
    \coordinate (v) at (1.5, 1.34);  

    \draw[thick, decorate, decoration={zigzag, segment length=4, amplitude=2}] 
        (u) arc[start angle=140, end angle=40, radius=1.9];

    \filldraw[black] (u) circle (2pt) node[above left] {\large $V$};
    \filldraw[black] (v) circle (2pt) node[above right] {\large $U$};

    \filldraw[red] (-0.45,1.97) circle (2pt) node[above left] {\large $W_2$};

    \filldraw[red] (0,0) circle (2pt);

    \draw[blue, thick, ->] 
        (-1.85, 1.23) arc[start angle=145, end angle=35, radius=2.25] -- 
        (1.55, 1.0) arc[start angle=35, end angle=145, radius=1.85] -- cycle;

    \node[blue] at (0.5,1.35) {\large $C_1$};

    \draw[->, blue, thick] (2.12, 2.12) arc[start angle=45, end angle=46, radius=3];

    \draw[->, blue, thick] (0.5, 1.73) arc[start angle=71, end angle=70, radius=2.1]; 

\end{tikzpicture}
\caption{The integration contour for the integral over $W_1$ in \eqref{eq: Snm W}. The integration is carried out along the unit circle in the complex plane, with the limits of integration being $U$ and $V$. The integrand has branch points at $W_{1,2} = U,V$ indicated by a black dot and a pole at $W_1=W_2$ indicated by a red dot. Depending on the value of $n$, the integrand can have a pole at the origin as well. The branch cut is taken along the unit circle, connecting $U$ and $V$ indicated by a serrated line. }
  \label{fig:contour}
  \end{figure}
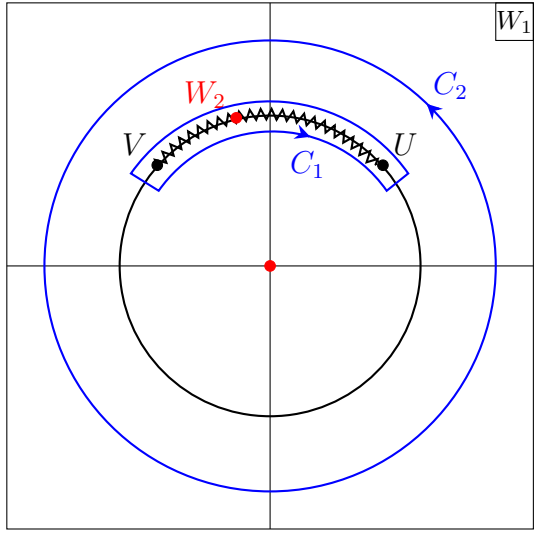
  
The integral in the above expression can be performed explicitly by mapping the cylinder to the complex plane via the conformal maps 
\begin{equation}
    W_1 = e^{i w_1}\,,\quad W_2 = e^{i w_2} \,, \quad U= e^{iw_u}\,, \quad V= e^{iw_v}\,. 
\end{equation}
In terms of these new coordinates, the matrix elements $S_{nm}$ are expressed as
\begin{equation}
    \label{eq: Snm W}
    S_{nm} = \int_U^V dW_1 dW_2\, G(W_1,W_2,U,V)\,.
\end{equation}
The integration is performed along the unit circle in the complex plane, with the integration limits being $U$ and $V$, as shown in figure~\ref{fig:contour}. The integrand has branch points at $W_{1,2} = U,V$ and a pole at $W_1=W_2$. Additionally, depending on the values of $n$ and $m$, there may be a pole at the origin. More precisely,
\begin{equation}
    G(W_1,W_2,U,V) = \frac{F(W_1,W_2,U,V)}{\left(U-W_1\right)^{3/2}\left(U-W_2\right)^{3/2}\left(V-W_1\right)^{3/2}\left(V-W_2\right)^{3/2}\left(W_1-W_2\right)^{4}}\,,
\end{equation}
where $F$ is meromorphic with a pole at $W_{1} =0$ for $n\leq-2$. Similarly, when $m\leq -2$, there is a pole at $W_2=0$.

To compute the integral in~\eqref{eq: Snm W}, we can exploit the analytic structure of the integrand using the residue theorem. We first focus on the $W_1$ integral. The function $G$ has branch points at $W_1=U,V$, a pole at $W_1=W_2$, and a pole at the origin when $n\leq -2$. We can place a branch cut along the unit circle connecting $U$ and $V$, as in figure~\ref{fig:contour}. Furthermore, the sign of $G$ changes as $W_1$ crosses this branch cut so that integrating $G$ along a thin counter-clockwise contour $C_1$ around the branch cut gives twice the value of the integral of $G$ from $U$ to $V$
\begin{equation}
    \int_U^V dW_1\, G(W_1,W_2,U,V) = \frac{1}{2} \int_{C_1} dW_1\, G(W_1,W_2,U,V)\,.
\end{equation}

Because $G(W_1,W_2,U,V)$ is meromorphic away from the branch cut, we can deform the contour $C_1$ to a large contour $C_2$  as shown in figure~\ref{fig:contour}, at the cost of including the (potential) residue at the origin
\begin{equation}
    \int_{C_1} dW_1\, G(W_1,W_2,U,V) = \int_{C_2} dW_1\, G(W_1,W_2,U,V) - 2 \pi i\, {\rm Res}_{W_1=0}G(W_1,W_2,U,V) \,.
\end{equation}
Lastly, the contour $C_2$ can be evaluated as we take the circle to infinity using the residue at infinity
\begin{equation}
    \int_{C_2} dW_1\, G(W_1,W_2,U,V) =  -2 \pi i\, {\rm Res}_{W_1=\infty}G(W_1,W_2,U,V)\,.
\end{equation}
We therefore find that the $W_1$ integral in~\eqref{eq: Snm W} is given by the residues of $G$ at the origin and infinity
\begin{equation}
    \int_U^V dW_1 G(W_1,W_2,U,V) = - \pi i \,  {\rm Res}_{W_1=0}G(W_1,W_2,U,V) -  \pi i \, {\rm Res}_{W_1=\infty}G(W_1,W_2,U,V)
\end{equation}

Let us denote this resulting expression as $H(W_2,U,V)$. We can apply a similar argument for the integral over $W_2$ which leads to
\begin{equation}
    \int_U^V dW_2 H(W_2,U,V) = - \pi i\, {\rm Res}_{W_2=0}H(W_2,U,V)  - \pi i\, {\rm Res}_{W_2=\infty}H(W_2,U,V) 
\end{equation}

Now, we have all the tools in order to compute $S_{nm}$ for any $n$ and $m$. Specifically, we find that $S_{nm}$ is symmetric and has the following structure
\begin{equation}
S = \pi^2(U-V)^4
\left[
\begin{array}{ccccccc}
     &   & \vdots & \vdots & \vdots &   &  \\
   \multicolumn{2}{c}{\smash{\raisebox{.5\normalbaselineskip}{$C(U,V)$}}}  &    &   &   &  \multicolumn{2}{c}{\smash{\raisebox{.5\normalbaselineskip}{$B(U,V)$}}} \\
     & \cdots & 0 & 0 & 0 &   &\cdots   \\
     & \cdots & 0 & 0 & 0 &   & \cdots  \\
     & \cdots & 0 & 0 & 0 &   & \cdots  \\
     &   &   &   &   &   &   \\
   \multicolumn{2}{c}{\smash{\raisebox{.5\normalbaselineskip}{$B^T(U,V)$}}} & \vdots & \vdots & \vdots &  \multicolumn{2}{c}{\smash{\raisebox{.5\normalbaselineskip}{$A(U,V)$}}} \\
\end{array}
\right]  +
\left[
\begin{array}{ccccccc}
     &   & \vdots & \vdots & \vdots &   &  \\
   \multicolumn{2}{c}{\smash{\raisebox{.5\normalbaselineskip}{$0$}}}  &    &   &   &  \multicolumn{2}{c}{\smash{\raisebox{.5\normalbaselineskip}{$D(U,V)$}}} \\
     & \cdots & 0 & 0 & 0 &   &\cdots   \\
     & \cdots & 0 & 0 & 0 &   & \cdots  \\
     & \cdots & 0 & 0 & 0 &   & \cdots  \\
     &   &   &   &   &   &   \\
   \multicolumn{2}{c}{\smash{\raisebox{.5\normalbaselineskip}{$D^T(U,V)$}}} & \vdots & \vdots & \vdots &  \multicolumn{2}{c}{\smash{\raisebox{.5\normalbaselineskip}{$0$}}} \\
\end{array}
\right] \,.
\end{equation}
Notably, if $n$ or $m$ take the values $-1$, $0$ or $1$, the matrix element $S_{nm}$ vanishes. This is because the corresponding gates $L_n$ or $L_m$ are global conformal transformations that leave the vacuum invariant.\footnote{This does not imply that the Bures metric is degenerate, as it is only a metric on the space of physically distinct states. The vanishing matrix elements for $-1 \leq n,m \leq 1$, which at first glance look like zero modes, determine the Bures metric along redundant directions in the space of states that appear only as a consequence of the particular parametrization we have chosen. The metric thus has to vanish along these directions.} We also note that there is the following symmetry
\begin{equation}\label{eq: S inversion}
    S_{nm}(U,V) = S_{(-n)(-m)}(U^{-1}, V^{-1})\,,
\end{equation}
because the $L_n$ generators are related to $L_{-n}$ by an inversion. To see this, recall that the $L_n$ are given by $z^{n+1}\partial_z$, and using $y=1/z$, we find
\begin{equation}
    L_n = z^{n+1} \partial_z = - y^{1-n} \partial_y = - I L_{-n} I\,,
\end{equation}
where $I$ is the inversion map.

The matrices $A,B$ and $D$ can be found explicitly. Specifically, we find that
\begin{align}
A(U,V)&= \small \begin{bmatrix}
 24  & 48  (U+V) & 10  \left(7 U^2+10 U V+7 V^2\right) & \cdots\\
 48(U+V) & 4 \left(25 U^2+46 U V+25 V^2\right) & 30  \left(5 U^3+11 U^2 V+11 U V^2+5 V^3\right) & \cdots \\
 \vdots &  \vdots & \frac{75}{16} \left(49 U^4+124 U^3 V+166 U^2 V^2+124 U V^3+49 V^4\right) & \cdots \\
 \vdots & \vdots & \vdots & \ddots
\end{bmatrix} 
\normalsize\,, \nonumber \\
    B(U,V)&=
\begin{bmatrix}
\vdots & \vdots & \vdots & \iddots \\
 -\frac{120 (U+V)}{U^{7/2} V^{7/2}} & -\frac{60  \left(3 U^2+10 U V+3 V^2\right)}{U^{7/2} V^{7/2}} & 0 & \cdots \\
 -\frac{96}{U^{5/2} V^{5/2}} &0 & -\frac{60  \left(3 U^2+10 U V+3 V^2\right)}{U^{5/2} V^{5/2}} & \cdots \\0  & -\frac{96 }{U^{3/2} V^{3/2}} & -\frac{120 (U+V)}{U^{3/2} V^{3/2}} & \cdots 
\end{bmatrix} \,,
    \\
    D(U,V) &= \small\begin{bmatrix}
\vdots & \vdots & \vdots  & \iddots \\ 
 0 & 0 & d_3(U,V)  & \cdots \\
 0 & d_2(U,V) &  0 & \cdots \\
d_1(U,V) & 0 &  0 & \cdots
\end{bmatrix}\normalsize\,, \nonumber \\
\end{align}
where the first few entries of the matrix $D(U,V)$ are given by
\begin{equation}
\begin{aligned}
    d_1(U,V)&=64 \pi ^2 \left(-\frac{(U+V) \left(U^2-10 U V+V^2\right)}{U^{3/2} V^{3/2}}-16\right)\,,\\
    d_2(U,V)&=-\frac{16 \pi ^2 \left(9 U^5-11 U^4 V-126 U^3 V^2+ 256 U^{5/2} V^{5/2} -126 U^2 V^3-11 U V^4+9 V^5\right)}{U^{5/2} V^{5/2}}\,,\\
    d_3(U,V)&=-\frac{5 \pi ^2 }{U^{7/2} V^{7/2}}\left(45 U^7-9 U^6 V-115 U^5 V^2-945 U^4 V^3 + 2048 U^{7/2} V^{7/2}\right.\\
    &\qquad\qquad\qquad\qquad\left.-945 U^3 V^4-115 U^2 V^5-9 U V^6+45 V^7\right)\,.
\end{aligned}
\end{equation}
The matrix $C$ is directly related to $A$ due to the property in~\eqref{eq: S inversion}.

Using the results obtained in this section, we have now derived the cost between infinitesimally close states $\rho(t)$ and $\rho(t+\delta t)$. As explained in the introduction \eqref{eq:complexity definition}, circuit complexity is defined as the minimal total cost between the initial and final state,
\begin{equation}\label{eq: total cost}
    C_\text{Bures}(t_i,t_f) = \text{min}\ \cF_\text{tot}(t_i,t_f) \quad \text{where} \quad \cF_\text{tot}(t) = \int_{t_i}^{t_f} dt' \cF_B(t')\,,
\end{equation}
where the minimum is taken over all possible choices of circuit.
This minimization procedure is difficult to perform in full generality.
However, a simple example of optimal transformations minimizing the total cost can be derived for small conformal transformations of the vacuum of the form \cite{Erdmenger:2021wzc}
\begin{equation}\label{eq:f pert}
  f(t,\phi+i t) = \phi + i t + i\epsilon t f_1(\phi)+\mathcal{O}\left(\epsilon^2\right)\,, 
\end{equation}
where $\epsilon$ is a small parameter and $f_1(\phi)$ is a periodic function.
These conformal transformations give the lowest total cost. Thus, the time integral of the Bures metric for \eqref{eq:f pert} determines the complexity of target states that are small perturbations of the vacuum.

In that case, the Fourier coefficients from \eqref{eq:f1 fourier} are given by $f^n = i\delta^{0n} + i\epsilon f_1^n + {\cal O}(\epsilon^2)$.
In CFT terms, the function $f_1(\phi)$ defines the final state of the full system after the perturbation. Up to a normalization factor, this final state can be written as
\begin{equation}\label{eq:Psi pert}
  \ket{\Psi(t)} = \ket{0} + \epsilon\, t \sum_{n=-\infty}^\infty f_1^n L_{-n}\ket{0}+\mathcal{O}\left(\epsilon^2\right)\,,
\end{equation}
where $L_n$ are the Virasoro generators. Notice that each Fourier coefficient $f_1^n $ controls the weight of the corresponding $L_{-n}$ mode in the final state. The subregion state is then found by tracing out the complement region
\begin{equation}
    \rho(t) = {\rm Tr}_{\bar{A}} |\Psi(t) \rangle \langle \Psi(t) |\,.
\end{equation}
Note that the coefficients of $S_{nm}$ with negative $n$ and $m$ do not always vanish. One might wonder why conformal transformations with non-vanishing $f^n$ for only negative $n$ would have non-zero Bures distance, as they don't change the state in eq.~\eqref{eq:Psi pert}. This apparent tension is resolved by the reality condition~\eqref{eq: reality}, which ensures that the only trivial conformal transformations are the global ones for which $n,m = -1,0,1$.

\section{Gravity dual to Bures metric and complexity}
\label{sec:Gravity dual Bures metric}

We now turn to the interpretation of the Bures metric in the dual gravity picture.
As reviewed in \autoref{sec: Setup}, the circuit built out of consecutive conformal transformations applied to the vacuum state has a dual interpretation as a particular time-slicing of the pure AdS$_3$ geometry \eqref{eq:bulk metric}.
The goal of this section is to identify an observable within this bulk geometry that is dual to the Bures metric between two states related by an infinitesimally small conformal transformation.

To do so, we generalize the analogous expression for the Fubini-Study metric derived in \cite{Erdmenger:2022lov}.
The basic idea is to rewrite the expression for the Bures metric in terms of the lengths of geodesics in the bulk geometry dual to a quantum circuit.
The geodesic length between two points $(z_1,\bar{z}_1)$ and $(z_2,\bar{z}_2)$ is given by
\begin{equation}
    l=l_{12}=\log\left(\frac{\sin\left((f(z_2)-f(z_1))/2\right)\sin\left((\bar{f}(\bar{z}_2)-\bar{f}(\bar{z}_1))/2\right)}{\epsilon_\text{UV}^2}\right)\,,
\end{equation}
where recall that the coordinates $z$ and $\bar{z}$ are introduced in~\eqref{eq: z zbar}.
We use this expression to define
\begin{equation}                   
\begin{aligned}
l_{1v}&=\log\left(\frac{\sin\left((w_1-w_v)/2\right)}{\epsilon_\text{UV}}\right)+ \text{antihol.~part}\,\\
l_{1u}&=\log\left(\frac{\sin\left((w_1-w_u)/2\right)}{\epsilon_\text{UV}}\right)+ \text{antihol.~part}\,\\
l_{2v}&=\log\left(\frac{\sin\left((w_2-w_v)/2\right)}{\epsilon_\text{UV}}\right)+ \text{antihol.~part}\,\\
l_{2u}&=\log\left(\frac{\sin\left((w_2-w_u)/2\right)}{\epsilon_\text{UV}}\right)+ \text{antihol.~part}\,\\
l_{uv}&=\log\left(\frac{\sin\left((w_u-w_v)/2\right)}{\epsilon_\text{UV}}\right)+ \text{antihol.~part}
    \end{aligned}
\end{equation}
such that the conformal cross ratio $x$ defined in~\eqref{eq: x equation} is given by 
\begin{equation}
    x
    =\sqrt{\frac{\partial^2_{z_u} l_{2u}\partial_{z_v}^2 l_{1v}}{\partial^2_{z_u} l_{1u}\partial_{z_v}^2 l_{2v}}}\,,\qquad \bar{x}
    =\sqrt{\frac{\partial^2_{\bar{z}_u} l_{2u}\partial_{\bar{z}_v}^2 l_{1v}}{\partial^2_{\bar{z}_u} l_{1u}\partial_{\bar{z}_v}^2 l_{2v}}}\,,
\end{equation}
and therefore the Bures metric~\eqref{eq: bures metric} can be written as follows
\begin{equation}
  \begin{aligned}
    \mathcal{F}_B = -\frac{c}{4}\int_u^v d\phi_1 d\phi_2\, \biggl[&\frac{(\partial_{z_u}\partial_{\phi_1} l_{1u})(\partial_{z_v}\partial_{t_1} l_{1v})(\partial_{z_u}\partial_{\phi_2} l_{2u})(\partial_{z_v}\partial_{t_2} l_{2v})}{\left(\partial_{z_u}\partial_{z_v}l_{uv}\right)^2}\frac{\sqrt{x} (1 - 9 x - 9 x^2 + x^3)}{16(x-1)^4}\\
    +&\frac{(\partial_{\bar{z}_u}\partial_{\phi_1} l_{1u})(\partial_{\bar{z}_v}\partial_{t_1} l_{1v})(\partial_{\bar{z}_u}\partial_{\phi_2} l_{2 u})(\partial_{\bar{z}_v}\partial_{t_2} l_{2v})}{\left(\partial_{\bar{z}_u}\partial_{\bar{z}_v}l_{uv}\right)^2}\frac{\sqrt{\bar{x}} (1 - 9 \bar{x} - 9 \bar{x}^2 + \bar{x}^3)}{16(\bar{x}-1)^4}\biggr].
  \end{aligned} \label{eq: Fbulk}
\end{equation}
Note that partial derivatives $\partial_{z_u}$ are derivatives w.r.t.~the holomorphic coordinate $z_u = u + i t_u$, keeping $\bar{z}_u = u - i t_u$ fixed. In the antiholomorphic part, partial derivatives $\partial_{\bar{z}_u},\partial_{\bar{z}_v}$ are used.
After all of the derivatives have been taken, we restrict to the constant time slice ($t_u = t_v = t_1 = t_2 = t$) on which the considered state is defined.

\begin{figure}
    \centering
    \begin{tikzpicture}
    \begin{scope}[shift={(-5,0)}] 
    \coordinate (A) at (0,0,2);
    \coordinate (B) at (2,0,0);
    \coordinate (C) at (-2,0,0);
    \coordinate (D) at (0,4,0);
    \coordinate (E) at (-1,2,0);
    \coordinate (F) at (1,2,0);
    \coordinate (G) at (0,0.7,0);
    \coordinate (H) at (0.45,1,0);
    \coordinate (I) at (0,3.3,0);
    \coordinate (J) at (0.85,1.5,0);
    \coordinate (K) at (-0.4,1.5,0);
    \coordinate (M) at (-0.25,2,0);
    \coordinate (N) at (0.75,2,0);
    \coordinate (Z) at (-0.75,2,0);
    \coordinate (L) at (0.25,2,0);

    \draw[thick, black] (B) -- (C) -- ++(0, 5, 0) -- ++(4, 0, 0) -- ++(0, -5, 0) -- cycle;
    \node[above right, black] at (-0.2,4,0) {\textbf{Boundary}};
    
    \draw[thick, black] (2,2,0) -- (-2,2,0)  -- ++(0, -4, -5) -- ++(4, 0, 0) -- ++(0, 4, 5) -- cycle;
    \node[below right, black] at (0,-1.5,-5.5) {\textbf{Bulk}};

    \draw[red, thick] 
        (E) .. controls (-0.7,2.1,0) .. (0,2,0) --
        (0,2,0) .. controls (0.7,1.9,0) .. (F);
    \node[above left, red] at (E) {\textbf{v}};
    \node[above right, red] at (F) {\textbf{u}};

    \draw[red, thick, dashed] 
        (E) .. controls (0,0.1,-1.5) .. (F);
    \fill[red!30, opacity=0.5] 
        (E) .. controls (0,0.1,-1.5) .. (F) -- cycle;

    \fill[blue!20, opacity=0.5] (G) -- (E) -- (F) -- cycle; 
    \draw[blue, thick] (G) -- (E); 
    \draw[blue, thick] (G) -- (F);
    \draw[blue, thick] (G) -- (H);
    \fill[blue!20, opacity=0.5] (I) -- (E) -- (F) -- cycle; 
    \draw[blue, thick] (I) -- (E); 
    \draw[blue, thick] (I) -- (F);
    \draw[blue, thick] (I) -- (H);
    \draw[blue, thick] (G) -- (J);
    \draw[blue, thick] (I) -- (J);
    \draw[blue, thick] (G) -- (K);
    \draw[blue, thick] (I) -- (K);

    \node at (0,-0.5) {(a)};
    
    \end{scope}

    \begin{scope}[shift={(5,0)}] 
    \coordinate (C) at (-2,0,0);
    \coordinate (F) at (0.12,2,0);
    \coordinate (Q) at (-2,3.5,0);
    \coordinate (P) at (-2,0.5,0);

    \draw[thick, black] (C) -- ++(0,0.5,0) ;
    \draw[thick, black] (-2,3.5,0) -- ++(0, 1.5, 0) ;
    \node[above right, black] at (-2.1,4,0) {\textbf{Boundary}};


    \node[above left, black] at (1,0.1,0) {\textbf{Bulk}};

    \draw[blue!40, opacity=1, thick] (Q) -- (P) ; 

    \fill[green!30, opacity=0.5] 
        (-2,2,0) .. controls (-1,1.5,0) .. (F)
        -- (F) .. controls (-1,2.5,0) .. (-2,2,0) -- cycle;

    \draw[red!30, opacity=1, thick] 
        (-2,2,0) .. controls (-1.5,2.1,0) .. (-1,2,0) --
        (-1,2,0) .. controls (-0.5,1.9,0) .. (F);
    
    \filldraw[red, opacity=1] (-2,2,0) circle (2pt);

    \draw[blue, thick] (Q) -- (F);
    \draw[blue, thick] (P) -- (F);
    \filldraw[blue] (F) circle (2pt);

    \node at (0,-0.5) {(b)};
    
    \end{scope}
    
\end{tikzpicture}
    \caption{Region of the bulk geometry probed by the Bures metric. \textbf{(a)}: The entanglement wedge (blue) of a boundary interval (solid red), showing the HRT surface (dashed red) and a Cauchy slice of the entanglement wedge (shaded red). \textbf{(b)}: Cross-section of (a). For an arbitrary spacelike interval, the union of geodesics connecting any two points on the interval sweeps a codimension-zero region of the bulk (shaded green).}
    \label{fig:bures metric bulk}
\end{figure}
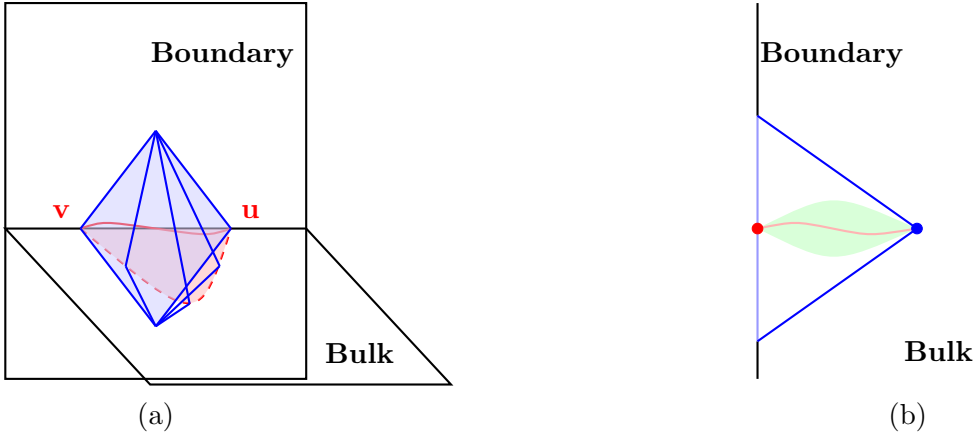

We have thus identified a purely geometric quantity dual to the Bures metric in the AdS$_3$ spacetime implementing the quantum circuit of interest. Together with~\eqref{eq: Bures metric}, this is the main result of this work. As expected for an observable characterizing feature of the reduced density matrix, $\mathcal{F}_B$ localizes to the entanglement wedge.
The geodesics, out of which $\mathcal{F}_B$ is built, stretch either between an interior point $\phi_1,\phi_2$ in the boundary interval and an interval endpoint or between the two endpoints $u$ and $v$.
Together, these geodesics sweep out a thin bulk codimension-zero surface contained in the entanglement wedge as illustrated in green on \autoref{fig:bures metric bulk} (b). Note that this codimension-zero surface anchors at a boundary codimension-one interval and the bulk codimension-two HRT surface. 

So far, we have dealt only with the bulk dual to the cost between infinitesimally close states, see eq.~\eqref{eq: total cost}.
From the geometric dual to the infinitesimal cost~\eqref{eq: Fbulk}, a geometric dual to the total cost for the entire circuit follows immediately, $\cF_\text{tot}(t_i,t_f) = \int_{t_i}^{t_f} dt' \mathcal{F}_B$.
This geometric dual is determined by a collection of geodesics sweeping out a codimension-zero region in the bulk, but this time anchoring on a boundary codimension-zero surface. This codimension-zero region in the bulk is localized in the union of all entanglement wedges for the interval $u \leq \phi \leq v$ within $t_i \leq t' \leq t_f$.

This notion of complexity is to be contrasted with other proposals for defining holographic subregion complexity which typically propose certain geometric quantities within a single entanglement wedge to be dual to some further unspecified measure of subregion complexity in the boundary theory.
In particular, it is often assumed that the reference state for these types of holographic complexity measures has no spatial entanglement \cite{Brown:2015lvg}.
The UV divergence inherent in such a definition of holographic complexity is then proposed to arise from creating the large amount of spatial entanglement in the vacuum state or any finite-energy excitation thereof \cite{Chapman:2017rqy,Jefferson:2017sdb}.
Here, the reference state is already the vacuum.\footnote{We can more generally take any descendant of the vacuum as well.}
Both reference and target state live on (boundary) Cauchy slices of the same bulk geometry.
Therefore, the complexity is dual to a bulk codimension-zero geometric object anchored between these two Cauchy slices.
The complexity is UV finite since we start with a highly entangled state in the first place.
Moreover, unlike the majority of previous work on holographic subregion complexity, the duality between the CFT complexity measure and its holographic equivalent presented here is derived from first principles, assuming only the standard AdS/CFT dictionary.

Finally, notice that because geodesics are Ryu-Takayanagi surfaces in AdS$_3$/CFT$_2$, there is also an interpretation of $\mathcal{F}_B$ in terms of a (complicated) combination of entanglement entropies.
This is reminiscent of the representation of holographic subregion complexity in terms of geodesic lengths derived using techniques from integral geometry in \cite{Abt:2017pmf,Abt:2018ywl}, although the two formulas of course differ in their details since they compute different quantities.\footnote{In \cite{Abt:2017pmf,Abt:2018ywl}, the geodesic length appears only quadratically while here we have a relation involving square roots and ratios of geodesic lengths. Moreover, in \cite{Abt:2017pmf,Abt:2018ywl} geodesics reaching outside the entanglement wedge appear which are counted only by the length of the part lying inside the entanglement wedge.}
We will have more to say about the relation between the Bures metric and holographic subregion complexity in the following section.

\section{Comparison with complexity = volume}
\label{sec:Holographic subregion complexity}

We now turn to holographic subregion complexity.
Previous research has established a connection between the Fubini-Study complexity of an infinitesimal conformal transformation of the vacuum and holographic complexity under the ``complexity=volume'' prescription. Specifically, it has been shown that the difference in volume between the maximal time slice of pure AdS$_3$, corresponding to the vacuum state $\ket{0}$, and the volume of the maximal time slice in a Ba\~nados geometry, the AdS$_3$ geometry dual to a CFT state $\ket{\Psi} = \ket{0} + \epsilon \ket{\Psi_1} + O(\epsilon^2)$ near the vacuum, is equal to the Fubini-Study complexity between $\ket{0}$ and $\ket{\Psi}$ up to third order in the perturbation parameter $\epsilon$ \cite{Erdmenger:2021wzc}.

In this section, we compute ``complexity=volume'' for subregions of both these Ba\~nados geometries and pure AdS$_3$, generalizing previous work of \cite{Flory:2018akz,Flory:2019kah,Flory:2020dja,Flory:2020eot,Erdmenger:2021wzc} to subsystems of the entire space. We then compare the differences in these complexities to the Bures metric between the respective reduced density matrices in the dual CFT. Unlike in the Fubini-Study case, the Bures metric generally differs from holographic subregion complexity at second order in perturbation theory, except when the subregion spans the entire space. In this case, the Bures metric coincides with the Fubini-Study metric. Nevertheless, the Bures metric exhibits notable similarities with holographic subregion complexity, suggesting that the true CFT dual may be closely related to the Bures metric -- at least for the small conformal transformations of the vacuum considered here.

Note that since we are comparing complexities of states that are perturbatively close, the first law of complexity framework~\cite{Bernamonti:2019zyy,Bernamonti:2020bcf} offers an alternative approach to the one presented in this section. This method is generalized and utilized in~\autoref{app: first law}. Note that this also allows for deriving perturbations to ``complexity=anything'' proposals.

\subsection{Complexity = volume for subregions}

The goal of this section is to derive the volume of a maximal Cauchy slice of the entanglement wedge of a single interval in a Ba\~nados geometry.
This quantity is also known as holographic subregion complexity, more specifically of the ``complexity=volume'' type \cite{Alishahiha:2015rta,Carmi:2016wjl,Ben-Ami:2016qex}.
Specifically, we consider codimension-one slices of the entanglement wedge that asymptote to an interval on a constant time slice of the Ba\~nados geometry on the boundary and intersect the RT surface.
We then maximize over the volumes of the codimension-one slices to get the holographic subregion ``complexity=volume''.

\paragraph{The bulk geometry}
Ba\~nados geometries are simply pure AdS$_3$ in a particular coordinate system \cite{Banados:1998gg}.
A constant time slice on the boundary of a Ba\~nados geometry maps to a diffeomorphism of a constant time slice of empty AdS$_3$.
In this section, we will work in Lorentzian signature such that the AdS$_3$ metric in global coordinates is given by
\begin{equation}
  ds^2 = -\cosh^2\rho\, d\tau^2 + d\rho^2 + \sinh^2\rho\, d\theta^2.
  \label{eq:global AdS3}
\end{equation}
Therefore the Cauchy slices of the entanglement wedge of a boundary interval on a constant time slice in the Ba\~nados geometry correspond to codimension-one slices in pure AdS$_3$ determined by $\tau(\rho,\theta)$ that asymptote to a spacelike interval on the boundary and that intersect the geodesic connecting the end-points of the spacelike interval.
To fix the conventions, we note that $\theta$ and $\tau$ are the real and imaginary parts of $w = f(t,z)$.

\paragraph{Geodesics in pure AdS$_3$}
The length of an arbitrary curve parametrized by $(\tau(\rho), \rho, \theta(\rho))$ from $\rho = \rho_*$ (the turning point of the geodesic in the bulk) to $\rho = \infty$ (the asymptotic boundary) is given by
\begin{equation}
  L = \int_{\rho_*}^\infty d\rho \sqrt{-\cosh^2\rho (\partial_\rho \tau)^2 + 1 + \sinh^2\rho (\partial_\rho \theta)^2}.
\end{equation}
To determine the geodesic,\footnote{We focus here on half of the geodesic: from the turning point up to the boundary.} we extremize this length functional by solving the Euler-Lagrange equations. We impose Dirichlet boundary conditions at the asymptotic boundary and Neumann-like boundary conditions at the turning point
\begin{equation}
    (\tau(\infty),\theta(\infty)) = (\tau_1 ,\theta_1 )\,, \quad (\tau'(\rho_*),\theta'(\rho_*))= (\infty,\infty)\,.
\end{equation}
Notably, the Neumann-like boundary conditions impose a single constraint, leaving one free parameter and the turning point $\rho_*$.

To fully determine the solution, we specify two endpoints at $(\tau_1,\theta_1) = (\bar \tau - \Delta \tau,\bar\theta - \Delta\theta)$, $(\tau_2,\theta_2) = (\bar \tau + \Delta \tau,\bar\theta + \Delta\theta)$. By symmetry of the configuration, the turning point is located at
\begin{equation}
   (\tau(\rho_*),\theta(\rho_*))=\left(\frac{\tau_1+\tau_2}{2},\frac{\theta_1+\theta_2}{2}\right)\,.
\end{equation}
With these constraints, the geodesic solution that extremizes the length is given by 
\begin{equation}
  \begin{aligned}
    \tau_\text{geo.}(\rho) &= \bar \tau - \frac{\pi}{4} + \frac{1}{2}\arctan\gamma(\Delta\theta,\Delta \tau),\\
    \theta_\text{geo.}(\rho) &= \bar \theta - \frac{\pi}{4} + \frac{1}{2}\arctan\delta(\Delta\theta,\Delta \tau),\\
  \end{aligned}
  \label{eq:geodesic-global-AdS3}
\end{equation}
where
\begin{equation}
  \begin{aligned}
    \gamma(\Delta\theta,\Delta\tau) &= \frac{\sin(2\Delta\tau)^2 + \cos(2\Delta\tau)\alpha\cosh^2\rho}{\sin(2\Delta\tau)\sqrt{-\sin^2(2\Delta\tau) - 2\cos(2\Delta\tau)\alpha\cosh^2\rho + \alpha^2\cosh^4\rho}}\,,\\
    \delta(\Delta\theta,\Delta\tau) &= \frac{\sin(2\Delta\theta)^2 + \cos(2\Delta\theta)\alpha\sinh^2\rho}{\sin(2\Delta\theta)\sqrt{-\sin^2(2\Delta\theta) - 2\cos(2\Delta\theta)\alpha\sinh^2\rho + \alpha^2\sinh^4\rho}}\,,\\
  \end{aligned}
\end{equation}
and $\alpha = \cos(2\Delta\tau) - \cos(2\Delta\theta)$. The radial position of the turning point is determined by
\begin{equation}
  \cosh^2\rho_* = \frac{2\cos^2(\Delta\tau)}{\cos(2\Delta\tau) - \cos(2\Delta\theta)}.
\end{equation}

\paragraph{Perturbative expansion}
To make the problem tractable, we consider time slices that are small perturbations of a constant time slice
\begin{equation}
  \tau(\rho,\theta) = \tau_0 + \epsilon \, \tau_1(\rho,\theta) + {\cal O}(\epsilon^2)\,,
  \label{eq:perturbation-expansion-maximal-volume-slice}
\end{equation}
where we will set $\tau_0=0$ for convenience in what follows. The boundary behavior of $\tau(\rho,\theta)$ defines the boundary time slice on which the state in the dual theory is defined. Hence, $\tau(\rho,\theta)$ is related to the conformal transformations discussed in~\autoref{sec:Bures metric CFT computation}
\begin{equation}
    \tau(\rho\to \infty,\theta) = \frac{1}{2i} \left(f(t_f,t_f+\theta) - \bar{f}(t_f,t_f-\theta) \right) - t_f\,,
\end{equation}
where recall $t_f$ denotes the time when the final state is reached, see eq.~\eqref{eq: total cost}. Correspondingly, the boundary asymptotics of $\tau_1(\rho,\theta)$ relate to the function $f_1(\theta)$ in eq.~\eqref{eq:f pert} and can be similarly expanded in Fourier modes\footnote{For simplicity, we apply the transformation only left moving coordinates here, $\bar{f}(t,\bar{z})=\bar{z}$.}
\begin{equation}
  \tau_1(\rho \to \infty,\theta) = \frac{1}{2}f_1(\theta) = \frac{1}{2}\sum_{n=-\infty}^\infty f_1^n e^{in\theta}.
\end{equation}
Recall that this transformation corresponds to a transformation of the state defined at the $\tau_0$ slice as in eq.~\eqref{eq:Psi pert}. Since we are expanding in $\epsilon$, the geodesic endpoints also need to be expanded in $\epsilon$,
\begin{equation}\label{eq: endpoints pert}
  \bar \tau = \epsilon\, \bar \tau_1+{\cal O}\left(\epsilon^2\right), \quad \Delta\tau = \epsilon\, \Delta\tau_1+{\cal O}\left(\epsilon^2\right)\,.
\end{equation}
Finally, we focus on a geodesic centered around $\bar\theta=0$, without loss of generality.

\paragraph{Determining the maximal volume slice}
Next, we want to find the location of the codimension-one slice that maximizes the volume
\begin{equation}
  V = \int d\rho d\theta \sqrt{\gamma}\,,
  \label{eq:volume}
\end{equation}
where the induced metric on an arbitrary codimension-one slice of pure AdS$_3$ is given by
\begin{equation}
  ds^2_\gamma = (1-\cosh^2\rho (\partial_\rho \tau)^2)d\rho^2 -2\cosh^2\rho (\partial_\rho \tau) (\partial_\theta \tau) d\rho d\phi + (\sinh^2\rho - \cosh^2\rho (\partial_\theta \tau)^2)d\theta^2.
  \label{eq:induced-metric}
\end{equation}
This is accomplished by solving the Euler-Lagrange equations for the volume functional. In what follows, we will proceed perturbatively in $\epsilon$ using the perturbation expansion \eqref{eq:perturbation-expansion-maximal-volume-slice}. The perturbative solution has been found in \cite{Erdmenger:2021wzc},
\begin{equation}
  \tau_1(\rho,\theta) =\sum_{n=-\infty}^\infty \tanh(\rho/2)^n \left(1+\frac{n}{\cosh\rho}\right)\left(C_1^n e^{in\theta} + C_2^{-n}e^{-in\theta}\right).
  \label{eq:maximal-volume-slice}
\end{equation}
The coefficients $C_1^n$ and $ C_2^n$ are fixed by appropriate boundary conditions. Specifically, we impose that
\begin{equation}
  \begin{aligned}
    \tau_1(\rho \to \infty,\theta) &= \frac{1}{2}f_1(\theta) \quad \text{(boundary asymptotics)},\\
    \tau_1(\rho_\text{geo.}(\theta),\theta) &= \tau_\text{geo.}(\theta) \quad \text{(intersection with RT surface)}.
  \end{aligned}
  \label{eq:boundary conditions}
\end{equation}
Here, $\rho_\text{geo.}(\theta)$ and $\tau_\text{geo.}(\theta)$ parametrize the location of the geodesic in terms of the angular coordinate $\theta$. These two functions are obtained by inverting \eqref{eq:geodesic-global-AdS3}.

\paragraph{Imposing boundary conditions}
The asymptotic boundary condition yields
\begin{equation}
  C_1^n + C_2^n = \frac{1}{2}f_1^n \quad \Leftrightarrow \quad C_1(\theta) + C_2(\theta) = \frac{1}{2}f_1(\theta)\,.
  \label{eq:asymptotic-bc}
\end{equation}
On the other hand, we are only approximately able to enforce that the maximal volume slice intersects the RT surface.\footnote{As explained in \autoref{app:FG expansion BC}, it is not always possible to impose the boundary condition in a near-boundary expansion. The reason is that the two sets of boundary conditions from \eqref{eq:boundary conditions} are incompatible for an arbitrarily curved boundary Cauchy slice at the interval endpoints on the asymptotic boundary in the strict $\rho=\infty$ limit. However, the boundary conditions are always compatible on a cutoff surface at finite $\rho$.} 
This approximation is achieved by truncating the sum in \eqref{eq:maximal-volume-slice} to finite order, i.e. $C_1^n,C_2^n=0$ for $|n|>n_\text{max}$.
By expanding the condition $\tau_1(\rho_\text{geo.}(\theta),\theta) = \tau_\text{geo.}(\theta)$ around the interval midpoint $\bar\theta$ and ensuring that the condition is satisfied at $\theta=\bar\theta$ and also that the first $2n_\text{max}$ derivatives match there
\begin{equation}\label{eq: approx maximal volume slice}
    \partial_\theta^n \tau_1(\rho_\text{geo.}(\theta),\theta) \big|_{\theta=\bar{\theta}}= \partial_\theta^n \tau_\text{geo.}(\theta)\big|_{\theta=\bar{\theta}}\,,\quad {\rm for} \ 0\leq n \leq 2 n_{\rm max}\,,
\end{equation}
we obtain a linear system of equations for 
$C_1^n$ and $C_2^n$. This approach fixes all coefficients simultaneously, and the value of each coefficient $C_{1,2}^n$ depends on $n_\text{max}$. As we increase $n_\text{max}$, the coefficients $C_{1,2}^n$ asymptote to a specific value for each $n$.

\paragraph{Divergent contributions to the volume}
Because the volume~\eqref{eq:volume} is UV divergent, we regularize it by introducing a UV cutoff. To facilitate subtracting off the vacuum contribution when comparing with the Bures metric, we choose the same UV cutoff as for the volume of a constant time slice of pure AdS$_3$.
In the Ba\~nados geometry, at the UV cutoff surface $\tilde\rho=\log(1/\epsilon_\text{UV})$, the induced metric
is given by
\begin{equation}
    ds^2_\text{cutoff}= \frac{1}{4\epsilon_\text{UV}^2}d\tilde x_+ d\tilde x_- + {\cal O}(1/\epsilon_\text{UV})\,,
\end{equation}
where $\tilde x_\pm = \phi \pm t$.
In terms of conformally transformed coordinates $\tilde x_+ = f_+(x_+)$ for $x_\pm = \theta \pm \tau$, we have
\begin{equation}
    ds^2_\text{cutoff} = \frac{1}{4\epsilon_\text{UV}^2}df_+( x_+)d x_- + {\cal O}(1/\varepsilon_\text{UV}) = \frac{1}{4\epsilon_\text{UV}^2}d x_+d x_-(1+\epsilon f_1'( x_+)) + {\cal O}(1/\epsilon_\text{UV})\,,
\end{equation}
where we have chosen to implement the conformal transformation on the left moving coordinate $x_+$.
We can absorb the $f_1$ dependence in these coordinates into a space and time-dependent prefactor
\begin{equation}
    ds^2_\text{cutoff} = \frac{1}{4\varepsilon_\text{UV}(x_+)^2}dx_+dx_- + {\cal O}(1/\epsilon_\text{UV})
\end{equation}
where
\begin{equation}
    \frac{1}{\varepsilon_\text{UV}(x_+)} = \frac{1}{\epsilon_\text{UV}}\left[1 + \frac{\epsilon}{2}f_1'(x_+) - \frac{\epsilon^2}{8}(f_1'(x_+))^2 + {\cal O}\left(\epsilon^3\right)\right]\,.
\end{equation}
In the coordinates $(x_+,x_-,\rho)$, the bulk metric is simply global AdS$_3$ \eqref{eq:global AdS3} with a UV cutoff surface at a spacetime dependent location $\rho=\log(1/\varepsilon_\text{UV}(\theta))$ in contrast to the constant UV cutoff surface in the Ba\~nados geometry $\tilde\rho=\log(1/\epsilon_\text{UV})$.

For holographic subregion complexity of an interval $[-\Delta\theta,\Delta\theta]$ with the cutoff prescription described above, the UV regulated volume is given by
\begin{equation}
    V = \int_{-\Delta\theta}^{\Delta\theta}d\theta \int_{\rho_\text{geo.}(\theta)}^{\log(1/\varepsilon_\text{UV})}d \rho \sqrt\gamma.
\end{equation}
For small perturbations \eqref{eq:perturbation-expansion-maximal-volume-slice} of the constant time slice, the determinant of the induced metric \eqref{eq:induced-metric} is given by
\begin{equation}
    \sqrt\gamma = \sinh\rho - \frac{\epsilon^2}{2} \frac{\cosh^2\rho\sinh^2\rho(\partial_\rho \tau_1)^2 + \cosh^2\rho(\partial_\theta \tau_1)^2}{\sinh\rho} + {\cal O}\left(\epsilon^3\right).
\end{equation}
Using the solution for $\tau_1(\rho,\theta)$ in~\eqref{eq:maximal-volume-slice}, we can evaluate the integral over $\rho$ to find
\begin{equation}\label{eq: regulated volume}
    \begin{aligned}
        V &= \int_{-\Delta\theta}^{\Delta\theta}d\theta\biggl(\cosh\rho- \frac{\epsilon^2}{2}\sum_{n_1,n_2=-\infty}^{\infty}\frac{n_1n_2}{\tanh(\rho/2)^{n_1+n_2}}\\
        & \hspace{2 cm} \times \biggl[
        \begin{aligned}[t]
            &-\cosh\rho(C_2^{n_1}+C_1^{n_1}\tanh(\rho/2)^{2n_1})(C_2^{n_2}+C_1^{n_2}\tanh(\rho/2)^{2n_2})\\
            &+\frac{(C_2^{n_1}-C_1^{n_1}\tanh(\rho/2)^{2n_1})(C_2^{n_2}-C_1^{n_2}\tanh(\rho/2)^{2n_2})}{\cosh\rho}\\
            &+2\frac{n_1n_2-1}{n_1-n_2}(C_1^{n_1}C_2^{n_2}\tanh(\rho/2)^{2n_1}-C_2^{n_1}C_1^{n_2}\tanh(\rho/2)^{2n_2})\biggr]\left.\biggr)\right._{\rho_\text{geo.}(\theta)}^{\log(1/\varepsilon_\text{UV})}\,,
        \end{aligned}
    \end{aligned}
\end{equation}
up to corrections of order ${\cal O}\left(\epsilon^3\right)$. To determine the UV divergent part, we note that $\left.\tanh(\rho/2)\right|_{\rho=\log(1/\varepsilon_\text{UV})} = {\cal O}(1)$ while $\left.1/\cosh(\rho)\right|_{\rho=\log(1/\varepsilon_\text{UV})} = {\cal O}(\varepsilon_\text{UV})$.
Thus, only the first term in the square brackets gives a UV divergent contribution. Hence, we find
\begin{equation}\label{eq: regulated volume final}
    V = \int_{-\Delta\theta}^{\Delta\theta}d\theta \int_{\rho_\text{geo.}(\theta)}^{\log(1/\varepsilon_\text{UV}(\theta))}\sqrt\gamma = \frac{1}{\epsilon_\text{UV}}\left(\Delta\theta + \frac{\epsilon}{4}(f_1(\Delta\theta)-f_1(-\Delta\theta)) + {\cal O} \left( \epsilon^2 \right)\right) + {\cal O}\left(\epsilon_\text{UV}^0\right).
\end{equation}
Notice that the UV divergent part of the regulated volume is given by the length of the interval measured w.r.t.~$ds^2_\text{cutoff}$.
If we keep the interval endpoints fixed at $\pm\Delta\theta$ in the $x_\pm$ coordinates, the length of the interval, and consequently also the UV divergent contribution to the regulated volume, changes slightly after we apply the conformal transformation. 

We aim to compute the complexity of a boundary interval on a constant time slice $\tau_0$ in Ba\~nados geometry dual to the state~\eqref{eq:Psi pert} and compare it to the complexity of the same boundary interval in pure AdS$_3$ dual to the vacuum. To this end, we consider intervals whose length doesn't change after the conformal transformation. This implies that we need to modify the location of endpoints 
\begin{equation}
    \pm \Delta\theta \to \Delta \theta_\pm=\pm\Delta\theta - \epsilon/2 f_1(\pm\Delta\theta) + {\cal O} \left(\epsilon^2\right)\,.
\end{equation}
The volume of these intervals is
\begin{equation}\label{eq: regulated volume shifted}
    V = \int_{\Delta\theta_-}^{\Delta\theta_+}d\theta \int_{\rho_\text{geo.}(\theta)}^{\log(1/\varepsilon_\text{UV}(\phi))}\sqrt\gamma = \frac{\Delta\theta}{\epsilon_\text{UV}} + {\cal O}\left(\epsilon_\text{UV}^0\right).
\end{equation}
The volume of the interval of the same length in pure AdS is found by setting $\epsilon=0$, and importantly has the same UV divergence.

\paragraph{Finite contributions to the volume}

At leading order in the perturbation expansion in $\epsilon$, the UV regulated volume~\eqref{eq: regulated volume} can be separated into two contributions
\begin{equation}
    V = \int_{\Delta\theta_-}^{\Delta\theta_+} d\theta \int_{\rho_\text{geo.}(\theta)}^{\log(1/\varepsilon_\text{UV}(\theta))}d\rho \left.\sqrt\gamma\right|_{\epsilon=0} + \int_{\Delta\theta_-}^{\Delta\theta_+} d\theta \int_{\left.\rho_\text{geo.}(\theta)\right|_{\Delta\tau=\bar \tau=0}}^{\log(1/\varepsilon_\text{UV}(\theta))}d\rho \left.\sqrt\gamma\right|_{{\cal O}(\epsilon^2)}+{\cal O}\left(\epsilon^3\right).
    \label{eq:volume finite part}
\end{equation}
The first term in this expansion integrates the ${\cal O}(1)$ part of $\sqrt\gamma$ while taking into account the shift of the RT surface under the conformal transformation up to quadratic order in $\epsilon$ as in~\eqref{eq: endpoints pert}. The second term integrates the ${\cal O}(\epsilon^2)$ part of $\sqrt\gamma$ up to the location of the unperturbed RT surface. Using
\begin{equation}
     \sinh^2\rho_\text{geo.}(\theta) =\frac{\sin ^2(2 \Delta \theta )}{(\cos (2 \Delta\tau)-\cos (2 \Delta \theta )) (\cos (2 \theta )-\cos (2 \Delta \theta ))}\,
\end{equation}
we find the first term is given by
\begin{equation}
    \int_{\Delta\theta_-}^{\Delta\theta_+} d\theta \int_{\rho_\text{geo.}(\theta)}^{\log(1/\varepsilon_\text{UV}(\theta))}d\rho \left.\sqrt\gamma\right|_{\epsilon=0} = {\cal O}\left(\frac{1}{\epsilon_\text{UV}}\right) + \pi + \epsilon^2 \Delta\tau_1^2 \pi \frac{\cos(\Delta\theta)}{\sin^2(\Delta\theta)}+{\cal O}\left(\epsilon^3\right)\,,
\end{equation}
where the UV divergent contribution is given in \eqref{eq: regulated volume shifted}. Note that setting $\epsilon=0$ reproduces the complexity of vacuum subregions.

The second term in \eqref{eq:volume finite part} depends on the boundary conditions \eqref{eq:boundary conditions} imposed on the RT surface which, as mentioned above, we are only able to impose approximately.
To evaluate this term, we use the approximate maximal volume slice whose derivation is explained around~\eqref{eq: approx maximal volume slice} to numerically compute that contribution to the volume. We therefore find an approximation to the volume $V_{\rm approx}$ for a given $n_{\rm max}$. By progressively improving the approximation of this contribution to the volume (increasing $n_\text{max}$), we consistently observe convergence of the volume according to a power-law scaling \begin{equation}
    V_\text{approx}(n_\text{max}) = V_\infty + b\, n_\text{max}^c\,,
\end{equation}
for high enough $n_\text{max}$ where $c < 0$. Leveraging this empirical relationship, we determine the holographic subregion complexity by calculating the volume at various $n_\text{max}$, fitting to the scaling law, and extracting the parameter $V_\infty$.

\subsection{Comparing the Bures metric complexity with subregion CV}\label{sec: comparison}
Using the numerical methods introduced in the last section, we now compare holographic subregion complexity with the Bures metric subregion complexity measure. Specifically, we compute the second-order perturbative corrections for both measures. We focus on the vacuum-subtracted holographic subregion complexity and the Bures metric complexity for various interval sizes, $\Delta\theta$, and for different choices of the small conformal transformation $f_1(\theta)$ from \eqref{eq:f pert} which defines the target state.

\begin{figure}
    \centering
    \includegraphics[width=0.75\linewidth]{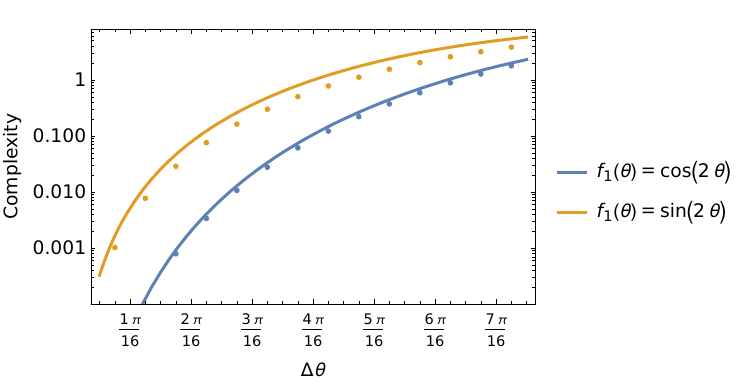}\\
    \includegraphics[width=0.75\linewidth]{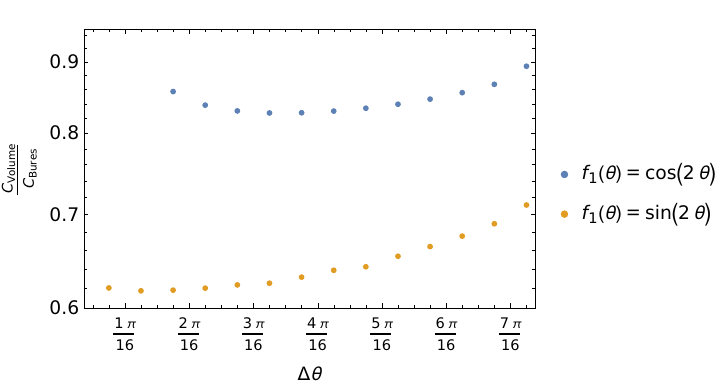}
    \caption{Top: Plot of the Bures metric complexity measure (solid line) and the vacuum subtracted holographic subregion complexity (dots) for different interval sizes. The increase in complexity with increasing interval size is very similar over several orders of magnitude (note the logarithmic scale), although the value of both quantities is not exactly the same. Bottom: Plot of the ratio of the two quantities shown on the top. The ratio is slowly varying with interval size but stays in a narrow range smaller than one.}
    \label{fig:HSC-BM-comparison1}
\end{figure}

Notably, the first-order perturbative contributions vanish for both measures, making the second-order terms the first non-trivial contributions to compare. For simplicity, we set  $\bar\theta=0$ without loss of generality, as shifting the interval midpoint corresponds to a shift in the argument of $f_1(\theta)$.

In all analyzed cases, we find no exact agreement between the two quantities. This is in contrast to the Fubini-Study complexity which agreed with ``complexity=volume'' for the entire space at the considered order in perturbation.
Specifically, we observe that holographic subregion complexity is consistently slightly smaller than the Bures metric complexity. The ratio between the two measures ranges between approximately $0.6$ and $0.9$ in the considered examples (see \autoref{fig:HSC-BM-comparison1}).
Despite the lack of exact agreement, the two quantities exhibit remarkably similar behavior as we vary the interval size and the conformal transformation that defines the target state. Both measures tend to zero as the interval size goes to zero, following a power law scaling $C \propto \Delta\theta^\lambda$ for $\Delta\theta \to 0$. Based on the available numerical data, we conclude that the exponent $\lambda$ in the power law is the same for holographic subregion complexity and the Bures metric complexity functional.

\begin{figure}
    \centering
    \includegraphics[width=0.75\linewidth]{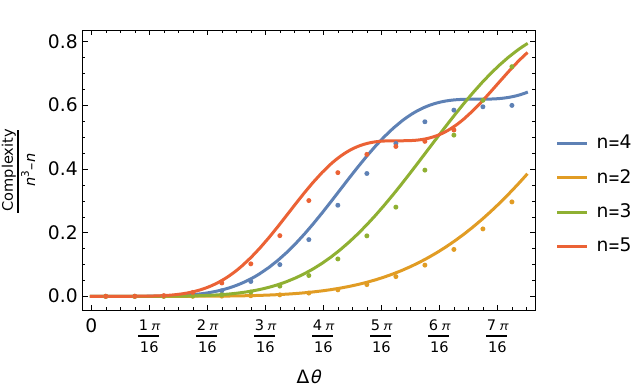}
    \caption{Plot of the Bures metric complexity measure (solid line) and the vacuum subtracted holographic subregion complexity (dots) for different interval sizes and different target states defined by $f_1(\theta) = \cos(n\theta)$. Higher values of $n$ lead to characteristic oscillations in the complexity that are visible both in the Bures metric and in the holographic complexity measure. Note that for plotting convenience, we have rescaled the complexity by $1/(n^3-n)$ to get the value for complexity into a similar range for all $n$.}
    \label{fig:HSC-BM-comparison2}
\end{figure}
Furthermore, we find striking similarities as we change the target state. 
We have computed both quantities for small conformal transformations of the form $f_1(\theta) = \cos(n\theta)$ for various values of $n$.
We find that both quantities increase as ${\cal O}(n^3)$ for fixed interval size. This scaling is consistent with the behavior of the Fubini-Study complexity, which is known to scale as $n^3-n$ for these target states.
Moreover, there are similarities between the behavior of both complexity measures as we vary the interval size. As we increase the size of the interval, both complexity measures increase monotonically for $n=2$, while on the other hand, there are oscillations for higher $n$.
Notably, both quantities show a very similar pattern of oscillations, as can be seen from \autoref{fig:HSC-BM-comparison2}.

In summary, while our numerical study is necessarily limited in scope due to the infinite range of potential target states and interval sizes, the findings indicate that vacuum-subtracted holographic subregion complexity and the Bures metric complexity functional exhibit remarkably similar behaviors. This similarity holds across various interval sizes and for target states representing small conformal transformations of the vacuum. Although these two quantities do not appear to be equal, their closely aligned behavior suggests a meaningful relationship between them.

\section{Discussion}
\label{sec:Discussion}

In this work, we have derived the Bures metric for mixed states associated with a single interval of CFT$_2$ on a circle, specifically for a class of descendant states of the vacuum. This result is given by the general formula (\ref{eq: Bures metric}). We have verified that in the pure state limit when the interval covers the entire spatial circle, the Bures metric correctly reduces to the Fubini-Study distance on the Hilbert space between the vacuum state and its conformal descendant. 

We have discussed that the Bures metric can serve as the cost function for defining the quantum circuit complexity of intervals in the CFT. The associated quantum circuit is realized by applying successive conformal transformations to the vacuum state. This quantum circuit can be holographically interpreted as the time evolution over a particular choice of time slicing in the AdS$_3$ spacetime. This allows us to identify the quantity (\ref{eq: Fbulk}) which directly computes the Bures metric from bulk data. This data consists of a family of geodesics connecting any two points of the interval, thus spanning a bulk codimension-zero subregion contained in the entanglement wedge of the interval, see~\autoref{fig:bures metric bulk}.\footnote{For intervals with very simple time profiles, the collection of geodesics can be a bulk codimension-one surface. In this case, this surface is a Cauchy surface of the entanglement wedge.} The bulk observable (\ref{eq: Fbulk}) then corresponds to an integral over this family of geodesics. In particular, the derivation of the equivalence between bulk and boundary descriptions of the Bures metric is a first-principles calculation assuming only the basic AdS/CFT dictionary.

Note also that the Nielsen complexity using the Bures metric as a cost function is intrinsically UV finite. From the perspective of the CFT computation, the finiteness of the Bures metric follows from its relation to the relative entropy (see eq.~\eqref{eq: bures metric}) which is a UV finite quantity. From the bulk computation, it might seem surprising that a geometric quantity in a bulk region that reaches up to the cutoff surface (see~\autoref{fig:bures metric bulk} (b))  is UV finite. However, the quantity in question~\eqref{eq: Fbulk} is defined in terms of changes in geodesic lengths with respect to variations of endpoints in the interval, which are UV finite. Since the Bures metric is finite, the finiteness of Nielsen complexity then follows from the choice of a reference state that is ``close'' to the target state, in the sense that we are not starting with a spatially unentangled reference state. This is in contrast to most previous works on complexity in the AdS/CFT context, 
in which a UV divergence of holographic complexity is associated to building up the large amount of entanglement in any finite energy QFT state \cite{Brown:2015bva,Susskind:2018pmk}.

We have also computed the holographic complexity of the CFT intervals using the subregion ``complexity = volume'' prescription for the same class of states. For general conformal transformations of the vacuum, we map the boundary interval in the associated Ba\~nados geometry to a boundary interval on empty AdS$_3$ with a spacetime dependent cutoff surface. This boundary interval lies along a nontrivial general time slice on the boundary. According to the general subregion ``complexity = volume'' prescription, the boundary conditions for the volume functional of codimension-$1$ surfaces are imposed on this generally nontrivial time slice, as well as on the codimension-$2$ RT surface anchored on the endpoints of the boundary interval. As explained in \autoref{app:FG expansion BC}, we find that these two sets of boundary conditions are in general not compatible when imposed at the exact asymptotic boundary. The qualitative reason is that since the time slice is arbitrarily curved, its curvature together with the continuity of the extremal surface spoil the gluing of the extremal surface to the RT surface near the anchor points. Therefore, the boundary conditions on the RT surface have been imposed perturbatively around the turning point of the RT surface, such that they remain valid everywhere up to the holographic cutoff surface. With this subtlety in mind, we used a numerical approach to impose the appropriate boundary conditions for the extremal volume slice and computed the subregion volume complexity for different shapes of boundary slices and sizes of the interval.

In~\autoref{sec: comparison}, we compared the Bures metric with holographic subregion ``complexity=volume''  by considering the difference in these two measures between subregions of the vacuum and target states of the form~\eqref{eq:Psi pert}, which consists of perturbations of the vacuum by Virasoro operators. We observed strong qualitative agreement in the behavior of both quantities. While they yield different quantitative results for the complexity of the quantum circuit under consideration, the two measures show striking similarities in their functional dependence on both interval size and the considered conformal transformations applied to the vacuum state suggesting a connection between the two definitions of complexity. Specifically,  we observed comparable behavior in scaling with respect to the Fourier index $n$ for fixed interval length, as well as similar patterns in size dependence for various Fourier modes, see~\autoref{fig:HSC-BM-comparison1} and \autoref{fig:HSC-BM-comparison2}. This contrasts with the quantitative match between ``complexity=volume'' and the Fubini-Study distance found in~\cite{Erdmenger:2021wzc}. In other words, while the Fubini-Study distance and the change in ``complexity=volume''  give the same leading order result for states of the form~\eqref{eq:Psi pert}, their natural generalizations for subregions -- namely the Bures metric and subregion ``complexity=volume'',  differ slightly.

Our quantitative comparison is facilitated by focusing on simple conformal transformations, which have a straightforward expansion in Fourier modes. Specifically, we choose transformations for which only a few Fourier modes are required. However, for more general conformal transformations, the sum over all non-trivial Fourier modes becomes complicated particularly due to the lack of a closed-form expression for the Fourier modes. Hence, performing this calculation for arbitrary transformations remains a challenge.

To conclude, we comment on some future directions. An interesting generalization of our results is the calculation of the Bures metric complexity for a mixed state of the CFT which is associated to a union of two disjoint intervals. An example in which this situation arises is a mixed state for two intervals for a pure state in a CFT. As an alternative example, one can start with a mixed state of a CFT, purify it by introducing a second copy, and then consider two intervals in the purification, with each interval belonging to a different copy of the CFT. These setups are interesting because it is well known that in holography the RT surface of the intervals experiences a phase transition \cite{Hayden:2011ag,Hartman:2013qma}. This phase transition in turn leads to a discontinuous transition in the volume of the entanglement wedge, causing holographic subregion complexity to experience a discontinuity \cite{Ben-Ami:2016qex,Abt:2017pmf,Abt:2018ywl,Craps:2022ahp}.\footnote{This discontinuity is expected to arise for any ``complexity = anything'' candidate} It would be interesting to diagnose this transition in the purely QFT framework of circuit complexity, using tools such as the Bures metric. A similar discontinuity is expected at finite temperature for holographic complexity on the entire space as the temperature crosses the Hawking-Page transition.

Another interesting direction is to extend the comparison of the Bures metric complexity with holographic subregion complexity to other complexity proposals, in particular within the ``complexity = anything'' family \cite{Belin:2021bga,Belin:2022xmt} adapted for subregions, using the methods from \autoref{app: first law}. The Bures metric complexity results could be used to identify a holographic proposal within the ``complexity=anything'' framework that would have better quantitative agreement with the QFT expectations for circuit complexity.

\section*{Acknowledgments}
We would like to thank Ben Craps, Michal P.~Heller, and Shan-Ming Ruan for useful discussions. Work at VUB was supported by FWO-Vlaanderen project G012222N and by the VUB Research Council through the Strategic Research Program High-Energy Physics. JH is supported by FWO-Vlaanderen through a Junior Postdoctoral Fellowship 12E8423N. MG is supported by FWO-Vlaanderen through a Junior Postdoctoral Fellowship 1238224N. The resources and services used in this work were provided by the VSC (Flemish Supercomputer Center), funded by the Research Foundation - Flanders (FWO) and the Flemish Government.

\begin{appendix}

\end{appendix}

\section{Reduction to Fubini-Study for pure states}\label{app: check}

In this appendix, we demonstrate how the Bures metric reduces to the Fubini-Study metric in the pure state limit, with particular attention to the analytic continuation in~\eqref{eq:fidelity-replica-trick}. We begin by examining the scenario where the analytic continuation to $m = n = 1/2$ is performed before taking the pure state limit. In this case, we verify that for mixed states, the term involving $\rho_2$ in~\eqref{eq: Bures derivative} vanishes when the result for $m, n > 1/2$ is analytically continued to $m = n = 1/2$. Subsequently, as discussed in \eqref{eq: reduce to FS}, the term associated with $\rho_1$ successfully yields the Fubini-Study metric when the pure state limit is applied after setting $m = n = 1/2$.

Conversely, if we reverse the order of limits by first taking the pure state limit and then continuing to $m, n \to 1/2$, we observe that the $\rho_2$ term does not vanish, as demonstrated below. Instead, this sequence of limits directly reproduces the Fubini-Study metric, while the $\rho_1$ term vanishes under these conditions.

We begin by considering mixed states and proceed to show that the $\rho_2$ term vanishes for such states. Specifically, we calculate $\Tr_A[\rho^{n(2m+1)-1}\rho_2]$ for $n, m > 1/2$, and subsequently take the limit $n = m = 1/2$, confirming that it correctly approaches zero.

To this end, we compute
\begin{equation}
  \begin{aligned}
    \Tr_A[\rho^{n(2m+1)-1}\rho_2] = &\frac{-1}{2}\int_{w_u}^{w_v} dw_1 dw_2 \sqrt{\det g(w_1)}\sqrt{\det g(w_2)}\\
    & \quad \times\ev{(T^t_t(t=0^-,w_1) - T^t_t(t=0^+,w_1))(T^t_t(t=0^-,w_2) - T^t_t(t=0^+,w_2))}\,,
  \end{aligned}
\end{equation}
on a Riemann surface with $n(2m+1)$ sheets and for which the two insertions labeled by $\phi_1$ and $\phi_2$ are on the same sheet. As explained in \autoref{fig:replica-trick}, we replace this expression with a two-point function on the complex plane via the uniformization map
\begin{equation}
  z = \left(\frac{\sin((w-w_u)/2)}{\sin((w-w_v)/2)}\right)^{1/(n(2m+1))}.
\end{equation}
Since $z(w=-i\epsilon) \sim e^{\frac{2\pi i}{n(2m+1)}}z(w=i\epsilon)$ for $\epsilon \to 0$, we get
\begin{equation}
  \begin{aligned}
    &\Tr_A[\rho^{n(2m+1)-1}\rho_2]\\
    & = \int d\phi_1 d\phi_2\, \partial_{\phi_1}f_1\partial_{t_1}f_1\partial_{\phi_2}f_2\partial_{t_2}f_2 \frac{c}{32n^4(2m+1)^4}\frac{\sin^4\bigl(\frac{w_u-w_v}{2}\bigr)}{\sin^2\bigl(\frac{w_1-w_u}{2}\bigr)\sin^2\bigl(\frac{w_1-w_v}{2}\bigr)\sin^2\bigl(\frac{w_2-w_u}{2}\bigr)\sin^2\bigl(\frac{w_2-w_v}{2}\bigr)}\\
    & \qquad \times \biggl[\left(x^{\frac{1}{2n(2m+1)}}e^{i\pi\left(\frac{1}{n(2m+1)}\right)}-x^{-\frac{1}{2n(2m+1)}}e^{-i\pi\left(\frac{1}{n(2m+1)}\right)}\right)^{-4} - \left(x^{\frac{1}{2n(2m+1)}}-x^{-\frac{1}{2n(2m+1)}}\right)^{-4}\biggr]\\
    & \quad + \text{antihol.~part}
    \end{aligned}
    \label{eq:result-Tr-rho2}
\end{equation}
For mixed states, $x \neq 1$, this expression vanishes as $n,m \to 1/2$. Therefore this term does not contribute to the Bures metric for the density matrix of a subregion, as claimed in the main text. If the pure state limit is then taken after this analytic continuation, it correctly reduces to the Fubini-Study metric, as discussed in \eqref{eq: reduce to FS}.

Alternatively, if we begin by taking the limit of the interval covering the entire space and then perform the analytic continuation $m,n\to 1/2$, the computation proceeds as follows.
For pure states on the entire space, we have $v=u+l$ with $l \to 2\pi$.
This implies
\begin{equation}
  \begin{aligned}
    x^\alpha - x^{-\alpha} & \to \alpha(l-2\pi)\left(\cot\left(\frac{w_u-w_2}{2}\right) - \cot\left(\frac{w_u-w_1}{2}\right)\right)\,,\\
    \sin^4\left(\frac{w_u-w_v}{2}\right) & \to \frac{(l-2\pi)^4}{16}\,.
  \end{aligned}
\end{equation}
Inserting this into \eqref{eq:result-Tr-rho2} gives
\begin{equation}
  \begin{aligned}
    \Tr_A[\rho^{n(2m+1)-1}\rho_2] \to & -\int d\phi_1 d\phi_2\, \partial_{\phi_1}f_1\partial_{t_1}f_1\partial_{\phi_2}f_2\partial_{t_2}f_2 \frac{c}{32} \sin\bigl(\frac{w_1-w_2}{2}\bigr)^{-4}\\
    &= -\bra{0}H^2\ket{0}\,,
  \end{aligned}
\end{equation}
which does not vanish. Indeed, the result is minus the Fubini-Study metric $\bra{0}H^2\ket{0} - \bra{0}H\ket{0}^2$ where $\bra{0}H\ket{0} = 0$, as expected from~\eqref{eq:Bures-metric}. Contrary to the case of mixed states, the contribution from the $\rho_1$ term in~\eqref{eq: Bures derivative} vanishes as it does not include any $x^\alpha - x^{-\alpha}$ denominators which could cancel with the vanishing $\sin^4\bigl(\frac{w_u-w_v}{2}\bigr)$ numerator in the $l \to 2\pi$ limit, see \eqref{eq:to continue}. Therefore, in this order of limits, it is instead the $\rho_2$ term that reduces to the Fubini-Study metric.

\section{First law of complexity for subregions}
\label{app: first law}

The comparison between the generalized Nielsen complexity using the Bures metric and subregion holographic ``complexity=volume'' in~\autoref{sec: comparison} relies on computing the difference in complexity between subregions of the vacuum, and subregions of a state that is perturbatively close to the vacuum. For this reason, we can readily use the framework of the first law of complexity~\cite{Bernamonti:2019zyy,Bernamonti:2020bcf} to present an alternative and more general approach to derive the results of~\autoref{sec: comparison}. In particular, this approach can be applied to an infinite class of holographic subregion complexity measures within the ``complexity=anything'' framework. To illustrate the method, we derive the change in holographic complexity between the states in subregions of the vacuum and subregions of the perturbed states~\eqref{eq:Psi pert} for the specific case of ``complexity=volume''.

\subsection{First law of circuit complexity}

Let us begin by summarizing the first law of complexity~\cite{Bernamonti:2019zyy,Bernamonti:2020bcf} which implies that to compute the change in complexity from an infinitesimal variation of the target state
\begin{equation}
    \delta {\cal C} = {\cal C}\left( \ket{\Psi_T+ \delta \Psi} \right) - {\cal C}\left( \ket{\Psi_T} \right)\,,
\end{equation}
one only needs information about the trajectory implemented by the circuit preparing the target state $\ket{\Psi_T}$ near the target state itself. More concretely, for a geometric (Nielsen) complexity model, one can associate a coordinate system $x^a$ to the space of unitary transformations or states, and the cost function induces a Lagrangian in terms of the coordinates and velocities
\begin{equation}\label{eq:C and F}
    {\cal C} = \min_{x(s)} \int_0 ^1 ds\, \cF(x^a(s),\dot{x}^a(s))\,,
\end{equation}
where the minimization is done over trajectories with fixed boundary conditions
\begin{equation}
    x^a(s=0) = x^a_0\,, \quad \quad  x^a(s=1) = x^a_1\,,
\end{equation}
which correspond to the coordinates of the reference and target unitary or state respectively. Because eq.~\eqref{eq:C and F} requires a minimization, the extremal trajectory $x^a_{\rm on-shell}(s)$ satisfies the Euler-Lagrange equations

\begin{equation}\label{eq:Euler}    \eval{\left(\frac{\partial \cF}{\partial x^a} - \frac{\partial}{\partial s} \frac{\partial \cF}{\partial \dot{x}^a}\right)}_{\rm on-shell}=0\,.
\end{equation}

Let us now vary the boundary conditions by allowing a one parameter family of reference and target states, parametrized by the coordinates $x^a_0(z)$ and $x^a_1(z)$ such that 
\begin{equation}
    x^a_0(z=0) = x^a_0\,, \quad \quad x^a_1(z=0) = x^a_1\,.
\end{equation}
Then as long as $|z|$ is kept small enough that none of the minimal trajectories connecting $x^a_0(z)$ and $x^a_1(z)$ pass a conjugate point, there is a smooth family of geodesics $x^a(a,z)$ minimizing~\eqref{eq:C and F} which satisfy the boundary conditions
\begin{equation}
    x^a(s=0,z) = x^a_0(z)\,, \quad \quad x^a(s=1,z) = x^a_1(z)\,.
\end{equation}

For small variations $\delta z$ around $z=0$ we have
\begin{equation}
    x^a(s,z) = x^a(s) + \delta x^a (s)
\end{equation}
where $x^a(s) \equiv x^a(s,z=0)$ and $\delta x^a(s) \equiv \delta z \eval{\partial_z x^a(s,z)}_{z=0} $. To leading order in $\delta z$, the change in complexity~\eqref{eq:C and F} from this change of boundary conditions is
\begin{equation}
\begin{aligned}
    \delta {\cal C}  & = \int_0^1 ds \left( \cF(x^a+\delta x^a, \dot{x}^a + \delta \dot{x}^a) - \cF(x^a,\dot{x}^a)\right) \\
    & = \eval{\frac{\partial \cF}{\partial \dot{x}^a} \delta x^a}_{s=0}^{s=1} + \int_0^1 ds \left(\frac{\partial \cF}{\partial x^a} - \frac{\partial}{\partial s} \frac{\partial \cF}{\partial \dot{x}^a}\right) \delta x^a + {\cal O}(\delta z^2)\,.
\end{aligned}
\end{equation}
The bulk term vanishes because the trajectory $x^a(s)$ connecting $x^a_0$ and $x^a_1$ satisfies the Euler-Lagrange equation~\eqref{eq:Euler}, so to linear order in $\delta z$, the increase in complexity is
\begin{equation}
    \delta {\cal C} = \eval{p_a \delta x^a}_{s=0}^{s=1}\,, \quad {\rm where} \quad p_a \equiv \frac{\partial \cF}{\partial \dot{x}^a}\,.
\end{equation}

To second order in $\delta z$, the increase in complexity is
\begin{equation}
\begin{aligned}
    \delta^{(2)} {\cal C} = \frac{1}{2}&\left(\frac{\partial^2 \cF}{\partial \dot{x}^a \partial x^b} \delta x^a \delta x^b + \frac{\partial^2 \cF}{\partial \dot{x}^a \partial \dot{x}^b} \delta x^a \delta \dot{x}^b \right)_{s=0}^{s=1} \\
    & + \frac{1}{2} \int_0^1 ds \left[\left(\frac{\partial^2 \cF}{\partial x^a \partial x^b} \delta x^b + \frac{\partial^2 \cF}{\partial x^a \partial \dot{x}^b}  \delta \dot{x}^b \right) -\frac{\partial}{\partial s}\left(\frac{\partial^2 \cF}{\partial \dot{x}^a \partial x^b}  \delta x^b + \frac{\partial^2 \cF}{\partial \dot{x}^a \partial \dot{x}^b}  \delta \dot{x}^b \right)\right]\delta x^a \,.
\end{aligned}
\end{equation}
The bulk term is equal to the variation of the Euler-Lagrange equation~\eqref{eq:Euler}. In other words, it corresponds to the difference of the Euler-Lagrange equation evaluated for the trajectory $x^a(s)+ \delta x^a(s)$ and $x^a(s)$. But both of these trajectories are geodesics minimizing~\eqref{eq:C and F}, so they each satisfy the Euler-Lagrange equation. The bulk term therefore vanishes and we are left with the boundary term, which corresponds to 
\begin{equation}
    \delta^{(2)} {\cal C} = \frac{1}{2} \delta p_a \delta x^a\,,\quad  {\rm where}\quad  \delta p_a = \delta \frac{\partial \cF}{\partial \dot{x}^a} = \frac{\partial^2 \cF}{\partial \dot{x}^a \partial x^b} \delta x^b + \frac{\partial^2 \cF}{\partial \dot{x}^a \partial \dot{x}^b}  \delta \dot{x}^b\,.
\end{equation}

To summarize, the change in complexity~\eqref{eq:C and F} from an infinitesimal change in reference and target state is 
\begin{equation}\label{eq:First law}
\delta {\cal C} = \eval{p_a \delta x^a}_{s=0}^{s=1} + \frac{1}{2} \eval{\delta p_a \delta x^a}_{s=0}^{s=1} + {\cal O} (\delta z^3)\,.
\end{equation}
Of course, we can keep the reference state fixed by restricting to the case $\eval{\delta x^a}_{s=0}=0$ which gives
\begin{equation}
\delta {\cal C} = \eval{p_a \delta x^a}_{s=1} + \frac{1}{2} \eval{\delta p_a \delta x^a}_{s=1} + {\cal O} (\delta z^3)\,.
\end{equation}

\subsection{First law of holographic complexity}

In the following, we generalize the approach of~\cite{Bernamonti:2019zyy,Bernamonti:2020bcf} 
previously used for circuit complexity to derive a first law of holographic complexity. Consider the holographic ``complexity=anything''  observables~\cite{Belin:2021bga,Belin:2022xmt} of codimension-one. In general, these are found by extremizing some geometric functional $\cF_1$ to specify a codimension-one surface $B$, and then evaluating a geometric functional $\cF_2$ (which can in principle be different than $\cF_1$) on the surface $B$. In the following, we will be restricting to holographic complexities for which both functionals are the same, \textit{i.e.,} $\cF_1=\cF_2$. That is, we are interested in the types of complexities
\begin{equation}\label{eq:Cgen}
    {\cal C}_{\rm gen} = \max_{\partial B = \Sigma}\left[ \frac{1}{G_N L} \int_B d^d\sigma \sqrt{h} \, \cF_2(g_{\mu\nu},\nabla_\mu;x^\mu(\sigma))\right]\,,
\end{equation}
where $x^\mu(\sigma)$ is the embedding of the codimension-one surface $B$, $\sigma^a$ are coordinates on $B$, and $\Sigma$ is the boundary surface on which the state is defined. The $\mu,\nu,\cdots$ indices run from $0$ to $d$, while the $a,b,\cdots$ indices run from $1$ to $d$. The function $\cF_2$ can in general depend on intrinsic and extrinsic quantities.

To connect with the first law of complexity~\eqref{eq:First law}, we rewrite~\eqref{eq:Cgen} as
\begin{equation}\label{eq:Cgen2}
    {\cal C}_{\rm gen} = \max_{\partial B = \Sigma} \int_B d^d\sigma \cF(x^\mu(\sigma),e^\mu_a(\sigma))\,,
\end{equation}
where 
\begin{equation}
    \cF(x^\mu(\sigma),e^\mu_a(\sigma)) = \frac{1}{G_N L} \sqrt{h} \, \cF_2(g_{\mu\nu},\nabla_\mu;x^\mu(\sigma))\,,
\end{equation}
and the $e^\mu_a \equiv \partial_a x^\mu$ dependence comes from the induced metric
\begin{equation}
    h_{ab} = e^\mu_a e^\nu_b g_{\mu\nu}\,.
\end{equation}

Extremization of~\eqref{eq:Cgen2} implies that the embedding $x^\mu(\sigma) $ satisfy the Euler-Lagrange equations on shell
\begin{equation}\label{eq:Euler2}
    \eval{\left(\frac{\partial \cF}{\partial x^\mu} - \frac{\partial}{\partial \sigma^a} \frac{\partial \cF}{\partial e^\mu_a}\right)}_{\rm on-shell} = 0\,.
\end{equation}
Consider now changing the state for which the holographic complexity~\eqref{eq:Cgen2} is evaluated by changing the boundary slice $\Sigma$ in which it is defined. This can be parametrized by a small change in the embedding 
\begin{equation}\label{eq:xprime}
    x'^\mu  = x^\mu + \delta x^\mu\,.
\end{equation}
The first order variation of holographic complexity from such a change in the boundary conditions is
\begin{equation}
    \delta {\cal C}_{\rm gen} = \int_{\partial B} d^{d-1}\sigma \tilde{m}_a p_\mu^a \delta x^\mu\,,
\end{equation}
where
\begin{equation}
    p_\mu^a \equiv \frac{\partial \cF}{\partial e^\mu_a}\,,
\end{equation}
and $\tilde{m}_a$ is the outwards pointing vector orthogonal to the boundary $\partial B$, normalized with respect to the metric $\eta_{ab}$. Similarly to the argument leading to the first law of circuit complexity, the bulk contribution vanishes because the surface $B$ is extremal and therefore the embedding satisfies the Euler-Lagrange equation. 

The second order change in complexity is
\begin{equation}
\begin{aligned}
    \delta^{(2)} {\cal C} = \frac{1}{2}&\int_{\partial B}d^{d-1}\sigma \left[\tilde{m}_a\left(\frac{\partial^2 \cF}{\partial e_a^\mu \partial x^\nu} \delta x^\nu + \frac{\partial^2 \cF}{\partial e_a^\mu \partial e_b^\nu} \delta e_b^\nu \right)\delta x^\mu\right] \\
    & + \frac{1}{2} \int_{B} d^d\sigma \left[\left(\frac{\partial^2 \cF}{\partial x^\mu \partial x^\nu} \delta x^\nu + \frac{\partial^2 \cF}{\partial x^\mu \partial e_a^\nu}  \delta e_a^\nu \right) -\frac{\partial}{\partial s}\left(\frac{\partial^2 \cF}{\partial e_a^\mu \partial x^\nu}  \delta x^\nu + \frac{\partial^2 \cF}{\partial e_a^\mu \partial e_b^\nu}  \delta e_b^\nu \right)\right]\delta x^\mu \,.
\end{aligned}
\end{equation}
The bulk term is the difference of the Euler-Lagrange equations evaluated for the embeddings $x^\mu+\delta x^\mu$ and $x^\mu$, but since both surfaces maximize~\eqref{eq:C and F}, the Euler-Lagrange equations vanish for each, and the only contribution left is the boundary term. 

We therefore find that up to second order, the change in complexity from a change in the boundary conditions is given by
    \begin{equation}\label{eq:First law holo}
    \delta {\cal C}_{\rm gen} = \int_{\partial B} d^{d-1}\sigma \tilde{m}_a \left( p_\mu^a  + \frac{1}{2} \delta p^a_\mu  \right) \delta x^\mu \,,
\end{equation}
where 
\begin{equation}
    \delta p_\mu^a = \frac{\partial p_\mu^a}{\partial x^\nu} \delta x^\nu+ \frac{\partial p_\mu^a}{\partial e^\nu_b} \delta e^\nu_b\,.
\end{equation}

\subsection{Complexity=volume}

Let us now apply the first law of holographic complexity to ``complexity=volume''. That is, we consider the function $\cF_2=1$, and therefore 
\begin{equation}
    \cF=\frac{\sqrt{h}}{G_N L} \,.
\end{equation}
The Euler-Lagrange equation is therefore
\begin{equation}\label{eq:K equation}
    0 = \frac{1}{\sqrt{h}} \partial_a \left(\sqrt{h} h^{ab} e_b ^\mu \right) + h^{ab} e^\nu_a e^\rho_b \Gamma^\mu_{\nu\rho} = K n^\mu \,,
\end{equation}
where $n^\mu$ is the unit normal vector to the surface $B$ and $K$ is the trace of the extrinsic curvature of $B$.  This equation states the familiar condition that $B$ has zero extrinsic curvature if it extremizes the volume functional.

For ``complexity=volume'' , the first order increase in complexity from a perturbation of the boundary slice~\eqref{eq:xprime} is
\begin{equation}\label{eq:deltaC1}
    \delta {\cal C}_{\rm gen} = \frac{1}{G_N L} \int_{\partial B} d^{d-1} \sigma \sqrt{\gamma} \, m^a e_a^\nu g_{\mu\nu} \delta x^\mu\,,
\end{equation}
where $\gamma$ is the determinant of the induced metric on $\partial B$, and $m^a$ is the outward pointing unit vector orthogonal to $\partial B$.

The second order increase in ``complexity=volume''  is 
\begin{equation}\label{eq:deltaC2}
    \delta {\cal C}_{\rm gen} = \frac{1}{4G_N L}\int_{\partial B} d^{d-1} \sigma \sqrt{\gamma} \, m_a \left(G^{a}_{\mu\nu} \delta x^\nu + G^{ab}_{\mu\nu} \delta e^\nu_b \right) \delta x^\mu\,,
\end{equation}
where
\begin{equation}
    \begin{aligned}
        G^a_{\mu\nu} &= \frac{2G_N L}{\sqrt{h}}  \frac{\partial^2 \cF}{\partial e^\mu_a \partial x^\nu} = \left( h^{ab} h^{cd} - 2 h^{ac} h^{bd}\right) e^\rho_b e^\sigma_c e^\tau_d g_{\mu \rho} \partial_\nu g_{\sigma\tau} + 2h^{ab} e^\rho_b \partial_\nu g_{\mu\rho} \,, \\
        G^{ab}_{\mu\nu} & = \frac{2G_N L}{\sqrt{h}}  \frac{\partial^2 \cF}{ \partial e^\mu_a \partial e^\nu_b } = 2\left( h^{ac} h^{bd} - 2 h^{ad} h^{bc}\right)  e^\rho_c e^\sigma_d g_{\mu \rho} g_{\nu\sigma}   + 2h^{ab} g_{\mu\nu} \,.
    \end{aligned}
\end{equation}
This can be written succinctly as
\begin{equation}
    \begin{aligned}
        G^a_{\mu\nu} &= \left( h^a _\mu h^{\sigma \tau} - 2 h^{a\sigma} h^{\tau}_\mu\right) \partial_\nu g_{\sigma\tau} + 2h^{a\rho}  \partial_\nu g_{\mu\rho} \,, \\
        G^{ab}_{\mu\nu} &= 2\left( h^{a}_\mu h^{b}_\nu - 2 h^{a}_\nu h^{b}_\mu\right) + 2h^{ab} g_{\mu\nu} \,,
    \end{aligned}
\end{equation}
where $h^{a\mu} = h^{ab}e^\mu_b$ and so on.

\subsection{Perturbations of the vacuum}

If we are interested in perturbations of the vacuum of a holographic CFT, the surface $\Sigma$ on the asymptotic boundary in which the state is defined is a constant time slice. Moreover, the holographic dual is empty AdS$_{d+1}$
\begin{equation}
    ds^2 = \frac{L^2}{z^2} \left(dz^2 + \eta_{ij} dx^i dx^j \right)\,,
\end{equation}
where the $i,j$ indices run from $0$ to $d-1$ and the asymptotic boundary is located at $z\to 0$. The metric $\eta = {\rm diag}(-1,1,\cdots,1)$ is the d-dimensional Minkowski metric. Since the boundary surface $\Sigma$ is a constant time slice, the maximal volume slice is simply given by $t=t_\Sigma$, so we can use $z$ and $x^i$ for $i\neq 0$ as the embedding coordinates of $B$
\begin{equation}
    \sigma^a = (\vec{x},z)\,, \quad x^\mu(\sigma) = (t_\Sigma,\vec{x},z)\,,
\end{equation}
and the deviation is captured by a function $\delta t(\vec{x},z)$
\begin{equation}\label{eq: variations dt}
    \delta x^\mu(\sigma) = (\delta t(\vec{x},z),\vec{0},0)\,.
\end{equation}
For these choices of coordinates, the einbeins and their variation are
\begin{equation}
    e^\mu_a = \delta^\mu_a\,, \quad \delta e^\mu_a = \delta^\mu_t \partial_a \delta t\,.
\end{equation}
It is also straightforward to compute
\begin{equation}
    h^{ab} = \frac{z^2}{L^2} \delta^{ab}\,,\quad h^{a\mu} = \frac{z^2}{L^2} \delta^{a\mu}\,, \quad  h^a_\mu = \delta^a_\mu
\end{equation}
as well as
\begin{equation}
    h^{\mu\nu} = \frac{z^2}{L^2} \delta^{\mu\nu}\,, \quad h^\mu_\nu = \delta^\mu_\nu\,, \quad {\rm for}\ \mu,\nu\neq 0\,,
\end{equation}
and zero otherwise. Additionally, the derivative of the metric is simply
\begin{equation}
    \partial_\mu g_{\nu\rho} = - \frac{2 L^2}{z^3} \eta_{\nu\rho} \delta_{\mu z}\,,
\end{equation}
where we have used a $d+1$-dimensional Minkowski metric $\eta_{\mu\nu}$. From this we find
\begin{equation}
    G^a_{\mu\nu} = - \frac{2d}{z} \delta^a_\mu \delta^z_\nu \,, \quad G^{ab}_{\mu\nu} =  2 \left( \delta^{a}_\mu \delta^{b}_\nu - 2 \delta^{a}_\nu \delta^{b}_\mu\right) + 2 \delta^{ab} \eta_{\mu\nu}.
\end{equation}
Lastly, we use that $\sqrt{\gamma} = \frac{L^{d-1}}{z^{d-1}}$ and $m^a = -\frac{z}{L}\delta^a_z$ to write
the first order increase in complexity~\eqref{eq:deltaC1} with respect to the vacuum
\begin{equation}\label{eq:deltaC1-vacuum}
\begin{aligned}
    \delta {\cal C}_{\rm gen} & = \frac{-1}{G_N L} \int_{\partial B} d^{d-1} \sigma \sqrt{\gamma} \,\frac{L}{z} \delta x^z\\ 
    & = \frac{-1}{G_N} \int_{\partial B} d^{d-1} \sigma \sqrt{\gamma} \frac{\delta z}{z} \\
    & = \frac{-L^{d-1}}{G_N \epsilon^{d}} \int_{\partial B} d^{d-1} \sigma \delta z \,.
\end{aligned}
\end{equation}

Similarly, the second order increase in ``complexity=volume'' ~\eqref{eq:deltaC2} with respect to the vacuum is 
\begin{equation}\label{eq:deltaC2-vacuum}
\begin{aligned}
    \delta {\cal C}_{\rm gen} &= \frac{-1}{4G_N L}\int_{\partial B} d^{d-1} \sigma \sqrt{\gamma} \, m_a \left(G^{a}_{\mu\nu} \delta x^\nu + G^{ab}_{\mu\nu} \delta e^\nu_b \right) \delta x^\mu\,\\
    & =\frac{-1}{4G_N L}\int_{\partial B} d^{d-1} \sigma \sqrt{\gamma} \, \frac{L}{z} \left(-\frac{2d}{z}  \delta x^z \delta^z_\mu + \delta^{zb} \eta_{\mu\nu} \delta^\nu_t \partial_b \delta t \right) \delta x^\mu\,\\
    & =\frac{1}{2G_N}\int_{\partial B} d^{d-1} \sigma \sqrt{\gamma} \, \frac{1}{z} \left(\frac{d}{z}  \delta z^2 - \delta t \partial_z \delta t \right)\,.
    \end{aligned}
\end{equation}

By plugging the embedding of the maximal volume surface $B$ into~\eqref{eq:deltaC2-vacuum}, one can reproduce the results of~\autoref{sec: comparison}. Similarly, one can use this method to reproduce the results for pure states in~\cite{Erdmenger:2021wzc}.

\section{Fefferman-Graham expansion}
\label{app:FG expansion BC}
In this appendix, we leverage the ambient metric formalism of Fefferman and Graham~\cite{Fefferman:2007rka} to express the increase in complexity given in eqs.~\eqref{eq:deltaC1} and \eqref{eq:deltaC2} 
in terms of geometric data of the boundary of the extremal codimension-$1$ slice $\partial B$ for a general asymptotically AdS$_{d+1}$ geometry. In doing so, we demonstrate why it is not always possible to impose the boundary conditions for subregion complexity=volume at the asymptotic boundary, as claimed in~\autoref{sec:Holographic subregion complexity}. This approach is particularly useful when considering the first law of holographic complexity for $d>2$. For $d=2$, the excited states can be related to a non-trivial time foliation of AdS$_3$. 
Therefore, this appendix additionally provides useful tools to guide any future work on holographic complexity in the higher dimensional setting.

\subsection{First Law}

We write the metric of any asymptotically AdS$_{d+1}$ spacetime as
\begin{equation}
    ds^2 = \frac{L^2}{z^2} \left( dz^2 + g_{ij}(z,x) dx^i dx^j\right)\,, 
\end{equation}
where the $i,j,\cdots$ indices run from $0$ to $d-1$, and the asymptotic boundary is located at $z\to 0$. For any solution to the bulk equations of motion, the metric $g_{ij}(z,x)$ can be expanded near the asymptotic boundary in terms of the boundary metric $\bar{g}_{ij}$ and stress tensor $\langle T_{ij}\rangle$
\begin{equation}
    g_{ij}(z,x^i) = \bar{g}_{ij} - \frac{z^2}{d-2}\left( \bar{R}_{ij} - \frac{\bar{R}}{2(d-1)} \bar{g}_{ij}\right) + \cdots + z^d \left( \tilde{g}_{ij} + f_{ij} \log \left( \frac{z}{L}\right)\right) + \cdots\,,\quad {\rm for}
    \ d>2\,,
\end{equation}
where $\bar{R}_{ij}$ is the boundary Ricci tenor and $\bar{R}$ is its trace. The term $\tilde{g}_{ij}$ is related to the boundary stress tensor $\langle T_{ij}\rangle$ and (derivatives of) the boundary metric $\bar{g}_{ij}$ as described in~\cite{Skenderis2000,deHaro:2000vlm}. The logarithmic terms appear only for even $d$, and their coefficient $f_{ij}$ is proportional to the metric variation of the conformal anomaly. The above equation is valid for $d>2$. For $d=2$, the Fefferman Graham expansion takes instead the form
\begin{equation}
g_{ij}( z, x^i) =  \bar{g}_{ij} +    z^2 \left(\tilde{g}_{ij}  + f_{ij} \log \left( \frac{z}{L} \right) \right)   + \cdots \,, \quad {\rm for}\ d=2\,.  
\end{equation}

Furthermore, it will be useful to make the following gauge choices:

\begin{equation}
    \sigma^a = (\sigma^m,z)\,, \quad h_{zm}=0 \,,
\end{equation}
where the $m,n,\cdots$ indices run from $1$ to $d-1$, and we have gauge fixed $\sigma^d=z$. The Euler-Lagrange equation~\eqref{eq:K equation} fixes the embedding functions near the asymptotic boundary as~\cite{Hernandez:2020nem}

\begin{equation}
\label{eq: xB}
x^i(z,\sigma^m) = \bar{x}^i(\sigma^m) + \frac{z^2}{2(d-1)} \bar{K} \bar{n}^i + \cdots \,, \quad {\rm for} \ d\geq 2\,,
\end{equation}
where $\bar{x}^i$ is the embedding of the boundary slice $\partial B=\Sigma$,  $\bar{K}$ is the trace of the extrinsic curvature of $\Sigma$ on the boundary, and $\bar{n}^i$ is the time-like unit normal to the boundary slice $\Sigma$, \textit{i.e.,} $\bar{g}_{ij} \bar{n}^i \bar{n}^j=-1$. 

Therefore in this gauge, the vielbein are, to leading order in $z/L$,
\begin{equation}
    e^z_a = \delta^z_a\,, \quad e^i_z = \frac{z}{d-1}\bar{K} \bar{n}^i\,, \quad e^i_{m} = \frac{\partial \bar{x}^i(\sigma^m)}{\partial \sigma^m}\,.
\end{equation}

The induced metric on $B$ can then be read off
\begin{equation}
    ds^2_B = h_{ab} d\sigma^a d\sigma^b = \frac{L^2}{z^2} \left[ \left(1 - \frac{z^2}{(d-1)^2} \bar{K}^2 \right)dz^2 + \bar{g}_{ij} \frac{\partial \bar{x}^i}{\partial \sigma^m} \frac{\partial \bar{x}^j}{\partial \sigma^n} d\sigma^m d\sigma^n \right] + \cdots\,.
\end{equation}
Then, using $m^a = -\frac{z}{L} \delta^a_z$, we can write the first order increase in complexity~\eqref{eq:deltaC1}
\begin{equation}
\begin{aligned}
    \delta {\cal C}_{\rm gen} & = \frac{-1}{G_N L} \int_{\partial B} d^{d-1} \sigma \sqrt{\gamma}  
\frac{z}{L} \left(e^z_z g_{zz} \delta z + e^j_z g_{ij} \delta x^i\right)\,,\\
 & = \frac{-1}{G_N} \int_{\partial B} d^{d-1} \sqrt{\gamma} \left(\frac{\delta z}{z} + \frac{\bar{K}}{d-1}  \bar{n}_i \delta x^i \right)\,,
 \end{aligned}
\end{equation}
which agrees with~\eqref{eq:deltaC1-vacuum} for subregions of the vacuum, for which$~\bar{K}=0$. To write the second order increase in complexity~\eqref{eq:deltaC2}, we also use that $\delta e^z_a=0$ and get
\begin{equation}
\begin{aligned}
    \delta {\cal C}_{\rm gen} & = \frac{-1}{4G_N L}\int_{\partial B} d^{d-1} \sigma \sqrt{\gamma} \, \frac{L}{z} \left(G^z_{zz} \delta z^2 + 2G^z_{(iz)} \delta x^z \delta x^i + G^z_{ij} \delta x^i \delta x^j + G^{z z}_{\mu i} \delta e^i_z \delta x^\mu + G^{z m}_{\mu i} \delta e^i_m \delta x^\mu\right)\,,\\
    & = \frac{1}{2G_N }\int_{\partial B} d^{d-1} \sigma \sqrt{\gamma} \left(\frac{d}{z^2} \delta z^2 + \frac{\Bar{K}}{z} \bar{n}_i \delta x^i \delta z  + \left( \frac{\bar{K}}{d-1}\right)^2 \left(\bar{n}_i \delta x^i \right)^2  - \frac{2}{d-1} \bar{K} \bar{n}_ i\delta z \partial_z \delta x^i\right.\\
    & \hspace{2 cm}\left.+ \frac{2}{z} \bar{g}_{ij}\delta x^i \partial_z \delta x^j + \frac{2}{z} \bar{h}^m_i \partial_m \delta x^i \delta z + \frac{2}{d-1}\left(\bar{n}_i \bar{h}^m_j \partial_m  -2 \bar{n}_j \bar{h}^m_i\ \right) \delta x^i \partial_m \delta x^j\right)\,,
\end{aligned}
\end{equation}
which is in agreement with~\eqref{eq:deltaC2-vacuum} in the limit of small perturbations from subregions of the vacuum, which imply $\bar{K}=0$, and in the parametrization~\eqref{eq: variations dt}.

\subsection{Subregion complexity}

The Fefferman-Graham expansion can also be used to show that the boundary conditions for subregion complexity=volume are not always possible to satisfy. Consider a boundary subregion $A$ and its HRT surface $\sigma_A$. A similar argument to the one leading to eq.~\eqref{eq: xB} leads to a FG expansion of the embedding $y^i(z,\sigma^m)$ of $\sigma_A$~\cite{Schwimmer:2008yh,Carmi:2016wjl,Chen:2020uac}
\begin{equation}
    \label{eq: xSigma}
    y^i(z,\sigma^m) = \bar{y}^i(\sigma^m) - \frac{z^2}{2(d-2)} \tilde{K}^i  + \cdots \,, \quad {\rm for} \ d>2\,,
\end{equation}
where $\tilde{K}^i$ is the trace of the extrinsic curvature of the separating surface $\partial A$. 
The subregion complexity=volume proposal requires that the maximal volume slice $B$ is anchored at the subregion $A$ and its HRT surface $\sigma_A$.
However, from eqs.~\eqref{eq: xB} and \eqref{eq: xSigma} it becomes clear that, unless the specific boundary slice $\Sigma$ and the separating surface $\partial A$ satisfy a very specific constraint
\begin{equation}
    (d-1) \tilde{K}^i \bar{n}_i = (d-2)\bar{K}\,, \quad {\rm for}\ d>2\,,
\end{equation} 
the asymptotic expansions of the HRT surface and the maximal volume surface $B$ are not compatible. 

For $d=2$, eq.~\eqref{eq: xB} is still valid, but the subleading term in the FG expansion of the HRT surface $\sigma_A$ is not constrained by $\tilde{K}$.\footnote{In $d=2$, the separating surface $\partial A$ consist of the two endpoints of the interval $A$, and their extrinsic curvature $\tilde{K}$ is simply zero.} Instead, it is related to the distance between the boundary endpoints as follows
\begin{equation}\label{eq: xSigma d=2}
    y^i_\pm(z) = \bar{y}^i_\pm \mp \frac{z^2}{ \Delta y^2}\Delta y^i + \cdots \,, \quad {\rm for}\ d=2\,,
\end{equation}
where $\bar{y}^i_\pm$ are the two endpoints of the interval $A$, $\Delta y^i=\bar{y}^i_+-\bar{y}^i_-$ and $\Delta y^2 = \Delta y^i \Delta y^j \bar{g}_{ij}$. Therefore in $d=2$, the asymptotic expansion of the maximal volume slice $B$~\eqref{eq: xB} and the HRT surface $\sigma_A$~\eqref{eq: xSigma d=2} are compatible only when
\begin{equation}
    \frac{2\Delta y{\cdot} \bar{n}}{\Delta y^2} = \pm \bar{K}_\pm\,,\quad {\rm for}\ d=2\,,
\end{equation}
where $\bar{K}_\pm$ is the extrinsic curvature of the boundary Cauchy slice $\Sigma$ at the endpoint $\bar{y}_\pm$.

\bibliographystyle{JHEP}
\bibliography{SubregionComplexity.bib}

\end{document}